\documentstyle{mn}
\newif\ifAMStwofonts
\AMStwofontstrue
\input epsfig.sty
\ifoldfss

  \ifCUPmtlplainloaded \else
    \NewTextAlphabet{textbfit} {cmbxti10} {}
    \NewTextAlphabet{textbfss} {cmssbx10} {}
    \NewMathAlphabet{mathbfit} {cmbxti10} {} 
    \NewMathAlphabet{mathbfss} {cmssbx10} {} 
  \fi
  \ifAMStwofonts
    \ifCUPmtlplainloaded \else
      \NewSymbolFont{upmath} {eurm10}
      \NewSymbolFont{AMSa} {msam10}
      \NewMathSymbol{\upi}     {0}{upmath}{19}
      \NewMathSymbol{\umu}     {0}{upmath}{16}
      \NewMathSymbol{\upartial}{0}{upmath}{40}
      \NewMathSymbol{\leqslant}{3}{AMSa}{36}
      \NewMathSymbol{\geqslant}{3}{AMSa}{3E}

      \let\geq=\geqslant 
    \fi
  \fi
\fi
\ifnfssone
  \newmathalphabet{\mathit}
  \addtoversion{normal}{\mathit}{cmr}{m}{it}
  \addtoversion{bold}{\mathit}{cmr}{bx}{it}

  \newmathalphabet{\mathbfit} 
  \addtoversion{normal}{\mathbfit}{cmr}{bx}{it}
  \addtoversion{bold}{\mathbfit}{cmr}{bx}{it}
  \newmathalphabet{\mathbfss} 
  \addtoversion{normal}{\mathbfss}{cmss}{bx}{n}
  \addtoversion{bold}{\mathbfss}{cmss}{bx}{n}
  \ifAMStwofonts
    \ifCUPmtlplainloaded \else
      \UseAMStwoboldmath
      \makeatletter
      \new@mathgroup\upmath@group
      \define@mathgroup\mv@normal\upmath@group{eur}{m}{n}
      \define@mathgroup\mv@bold\upmath@group{eur}{b}{n}
      \edef\UPM{\hexnumber\upmath@group}
      \new@mathgroup\amsa@group
      \define@mathgroup\mv@normal\amsa@group{msa}{m}{n}
      \define@mathgroup\mv@bold\amsa@group{msa}{m}{n}
      \edef\AMSa{\hexnumber\amsa@group}
      \makeatother
      \mathchardef\upi="0\UPM19
      \mathchardef\umu="0\UPM16
      \mathchardef\upartial="0\UPM40
      \mathchardef\leqslant="3\AMSa36
      \mathchardef\geqslant="3\AMSa3E

      \let\geq=\geqslant 
    \fi
  \fi
\fi
\ifnfsstwo

  \DeclareMathAlphabet{\mathbfit}{OT1}{cmr}{bx}{it}
  \SetMathAlphabet\mathbfit{bold}{OT1}{cmr}{bx}{it}
  \DeclareMathAlphabet{\mathbfss}{OT1}{cmss}{bx}{n}
  \SetMathAlphabet\mathbfss{bold}{OT1}{cmss}{bx}{n}
  \ifAMStwofonts
    \ifCUPmtlplainloaded \else
      \DeclareSymbolFont{UPM}{U}{eur}{m}{n}
      \SetSymbolFont{UPM}{bold}{U}{eur}{b}{n}
      \DeclareSymbolFont{AMSa}{U}{msa}{m}{n}
      \DeclareMathSymbol{\upi}{0}{UPM}{"19}
      \DeclareMathSymbol{\umu}{0}{UPM}{"16}
      \DeclareMathSymbol{\upartial}{0}{UPM}{"40}
      \DeclareMathSymbol{\leqslant}{3}{AMSa}{"36}
      \DeclareMathSymbol{\geqslant}{3}{AMSa}{"3E}

      \let\geq=\geqslant 
    \fi
  \fi
\fi
\ifCUPmtlplainloaded \else
  \ifAMStwofonts \else 
    \def\upi{\pi}
    \def\umu{\mu}
    \def\upartial{\partial}
  \fi
\fi

\title{Systematic Construction of Exact MHD models for 
Astrophysical Winds and Jets}
\author[N. ~Vlahakis \& K. ~Tsinganos]
       {N.~Vlahakis$^1$, and K.~Tsinganos$^{1,2}$ \\
 $^1$Department of Physics, University of Crete, GR-710 03 Heraklion, Crete,
 GREECE\\
 $^2$Foundation for Research and Technology Hellas (FORTH),
 GR-711 10 Heraklion, Crete, GREECE\\
 vlahakis@physics.uch.gr, tsingan@physics.uch.gr
 }
 \date{Submitted: ?}
 \date{Accepted ?; Received ?}

\pagerange{\pageref{firstpage}--\pageref{lastpage}}
\pubyear{1994}

\begin{document}

\maketitle

\label{firstpage}

\begin{abstract}
By a systematic method we construct general classes of exact
and selfconsistent axisymmetric
MHD solutions describing flows which originate at the near environment of a
central gravitating astrophysical object.
The unifying scheme contains two large groups of exact MHD outflow models,
(I) meridionally self-similar ones with {\it spherical} critical surfaces and
(II) radially self-similar models with {\it conical} critical surfaces.
The classification includes known polytropic models, such as the classical Parker
model of a stellar wind and the Blandford and Payne (1982)
model of a disk-wind; it also contains nonpolytropic models, such as those
of winds/jets in Sauty and Tsinganos (1994), Lima et al (1996) and
Trussoni et al (1997).
Besides the unification of these known cases under a common scheme,
several new classes emerge and some are briefly
analysed; they could be explored for a further understanding of the physical
properties of MHD outflows from various magnetized and rotating astrophysical
objects in stellar or galactic systems.
\end{abstract}


\section{Introduction}

A widespread phenomenon is astrophysics is the outflow of plasma from the 
environment of stellar or galactic objects, either in the form of a noncollimated 
wind (Parker 1958, Feldman et al 1996), or, in the form of collimated jets 
(Blandford \& Rees 1974, Biretta 1996, Ferrari et al, 1996). 
These outflows not only occur around typical stars and the nuclei of 
many radio galaxies and quasars, but they are also associated with young 
stars, older mass losing stars and planetary nebulae, symbiotic stars, 
black hole X-ray transients, low- and high-mass X-ray binaries and cataclysmic 
variables (for recent reviews see {\it e.g.}, Ray 1996, Kafatos 1996, 
Mirabel \& Rodriguez 1996, Livio 1997). 
Even for the two spectacular rings seen with the HST in SN1987A, it has been 
proposed that they may be inscribed by two precessing jets from an object 
similar to SS433 on a hourglass-shaped cavity which is created by 
nonuniform winds of the progenitor star (Burderi and King, 1995, 
Burrows et al 1995). Also recently, in the well known long jet of the 
distant radio galaxy NGC 6251 
an about $10^3$ light-year-wide warped dust disk perpendicular to the 
main jet's axis has been observed by HST to surround and reflect UV light 
from the bright core of the galaxy which probably hosts a black hole 
(Crane \& Vernet 1997).

Nevertheless, despite their abundance the questions of the formation, 
acceleration and propagation of nonuniform winds and jets have not 
been fully resolved. One of the main difficulties in dealing with the 
theoretical problem posed by cosmical outflows is that their dynamics 
needs to be described - even to lowest order - by the highly intractable 
set of the MHD equations. As is well known, this is a nonlinear system 
of partial differential equations with several critical points, etc, and 
only very few classes of solutions are available for axisymmetric systems 
obtained by assuming a separation of variables in several key 
functions. This 
hypothesis allows an analysis in a 2-D geometry of the full MHD equations 
which reduce then to a system of ordinary differential equations. 
The basis of the self-similarity treatment is the prescription of a scaling law 
in the variables as a function of one of the coordinates. The choice of the
scaling variable depends on the specific astrophysical problem. 

In spherical coordinates ($r\,, \theta\,, \phi$), a {\it first} broad 
class for describing outflows are the so-called meridionally 
self-similar MHD models. Parker's (1958) classical modeling 
of the spherically symmetric polytropic solar wind is the simplest member of 
this class.
A new class of such type of models for describing magnetized 
and rotating MHD outflows from a central gravitating object has also been 
examined (Sauty \& Tsinganos 1994 (henceforth ST94), 
Lima et al, 1996, Trussoni et al 1997).  For example, an energetic criterion 
for the transition of an asymptotically conical outflow from an inefficient 
magnetic rotator to an asymptotically cylindrical outflow from an efficient 
magnetic rotator was derived.  In the present paper, it will be shown  
that this special class of meridionally self-similar solutions is one 
of the simplest possible meridionally self-similar models. Furthermore, a new 
interesting member of this class of 
radially self-similar MHD models will be briefly 
sketched. 

A {\it second} broad class of solutions contains the radially 
self-similar MHD models. Bardeen \& Berger (1978) presented the 
first such models in the context of hydrodynamic and polytropic galactic 
winds. Nevertheless, their generalisation to a cold magnetized plasma 
by Blandford \& Payne 1982 (henceforth BP82), remains widely known because 
of their success 
in showing for the first time that astrophysical jets can be accelerated 
magnetocentrifugally from a Keplerian accretion disk, {\it if} the poloidal 
fieldlines are inclined by an angle of 60$^o$, or less, to the disk 
midplane (but see also, Cao 1997).  
A further extension has been presented by Contopoulos \& 
Lovelace (1994) for a hot plasma with a more general parametrization of 
the magnetic flux on the disc, while these models form the basis of 
several investigations of accretion-ejection flows from stars and AGN 
(Konigl 1989; Ferreira \& Pelletier 1995; Ferreira 1997; 
Li 1995, 1996). In this paper it will be shown 
that this special class of radially self-similar solutions is one of the  
simplest possible such models. Furthermore, a new 
interesting member of the radially self-similar MHD models will be sketched. 
 
In subsection 2.1 we use a simple theorem in order to construct several 
classes of meridionally selfsimilar solutions and the resulting cases 
are then summarized in Tables 1 and 2. The general method is next applied in 
subsection 2.2 to a step by step construction of a new model for collimated 
outflows which is also briefly sketched there. 
In section 3 the other remaining possibility in spherical coordinates,  
i.e., radial self similarity is taken up.  The 
resulting cases are summarized in Table 3 while a new model is also 
briefly sketched which gives asymptotically cylindrical, paraboloidal 
and conical streamlines. Finally, the results are summarized in Sec. 4.

\section[]{Meridionally selfsimilar \\* MHD outflows}

Consider the steady (\(\partial/\partial t=0\)) hydromagnetic equations. 
They consist of a set of eight coupled, nonlinear, partial differential 
equations expressing momentum, magnetic and mass flux conservation,  
together with Faraday's law of induction in the ideal MHD limit, 

\begin{equation}\label{momentum}
\rho \left( \vec{V}\cdot\vec{\nabla}\right)\vec{V}=
\frac{\left(\vec{\nabla}\times
\vec{B}\right)\times\vec{B} } {4\pi}
-\vec{\nabla}P-\rho\vec{\nabla}{\cal V}
\,,
\end{equation}
\begin{equation}\label{fluxes+F} 
\vec{\nabla}\cdot\vec{B}=0\,,\quad \vec{\nabla}\cdot\left(\rho\vec{V}\right)=0
\,,\quad \vec{\nabla}\times\left(\vec{V}\times\vec{B}\right)=0
\,.
\end{equation}
\noindent
\(\vec{B}\), \(\vec{V}\), \(-\vec{\nabla}{\cal V}=
-\vec{\nabla} \left( -{ {\cal GM}}/ {r} \right)\) denote the magnetic, 
velocity and external gravity fields, respectively, while \( \rho\) 
and $P$ the gas density and pressure.
With axisymmetry $(\partial/\partial \phi=0)$, we may introduce the magnetic 
flux function $A$, such that three free integrals exist for the total specific 
angular momentum carried by the flow and the magnetic field, $L(A)$, the 
corotation angular velocity of each streamline at the base of the flow,  
$\Omega(A)$ and the ratio of the mass and magnetic fluxes, $\Psi_A(A)$
(Tsinganos 1982). In terms of these integrals and the square of 
the poloidal Alvf\'en Mach number (or Alfv\'en number),

\begin{equation}\label{Alfven}
M_{}^{2}=\frac{4 \pi\rho V_{p}^{2}}{B_{p}^{2}}=\frac {\Psi_{A}^{2}}{4 \pi 
\rho}\,, 
\end{equation}
the magnetic field and bulk flow speed are given in spherical coordinates 
$(r,\theta,\phi)$ 
\begin{equation}\label{Bfield}
\vec{B}=\vec{\nabla}  \times \frac {A (r, \theta) \hat {\phi}} {r\sin \theta } 
-\frac{L\Psi_{A}-r^{2}\sin ^{2}\theta \Omega \Psi_{A}}{r\sin \theta  
(1-M_{}^{2})}\hat {\phi}
\,,
\end{equation}
\begin{equation}\label{Vfield}
\vec{V} =\frac{\Psi_{A}} {4 \pi \rho} \vec{\nabla}\times \frac{A(r, \theta)
\hat {\phi}}{r\sin \theta }+\frac{r^{2}\sin ^{2}\theta \Omega-L M_{}^{2}}{r\sin 
\theta  (1-M_{}^{2})}\hat {\phi}
\,.
\end{equation}

\noindent 
To construct classes of exact solutions, we shall make two crucial 
assumptions:
\begin{enumerate}
  \item that the Alfv\'en number $M_{}^{}$ is some function of the
dimensionless radial distance $R={r/ r_{\star}}$,
\begin{equation}\label{assumptions1}
M_{}^{} = M(R)\,,
\end{equation}
and
\\
  \item that the poloidal velocity and magnetic fields have a dipolar
angular dependence,
\begin{equation}\label{assumptions2}
A= {r_{\star}^2 B_{\star} \over 2}{\cal A}\left(\alpha\right)\,,
\qquad
\alpha=\frac{R^2}{G^2\left(R\right)}\sin^2 \theta
\,.
\end{equation}
\end{enumerate}
By choosing $G\left(R=1\right)=1$ at the Alfv\'en transition $R=1$, 
$G(R)$ evidently measures the cylindrical distance $\varpi$ to the 
polar axis of each fieldline labeled by $\alpha$, normalized to its 
cylindrical distance $\varpi_{\alpha}$ at the Alfv\'en point, 
$G\left(R\right)={\varpi}/{\varpi_{\alpha}}$. For a smooth crossing of 
the Alfv\'en sphere $R=1$ [$r=r_{\star}, \theta = \theta_a (\alpha) $], 
the free integrals $L$ and $\Omega$ are related by 
\begin{equation}\label{regularity_phi}
{L\over \Omega } = 
\varpi_{\alpha}^2 (A) = r_{\star}^2 \sin^2 \theta_a (\alpha) =r_{\star}^2 
\alpha \,.
\end{equation}
Therefore, the second assumption is equivalent with the statement that  
at the Alfv\'en surface the cylindrical distance $\varpi_{\alpha}$ of each 
magnetic flux surface $\alpha=const$ is simply proportional to 
$\sqrt{\alpha}$. 
\\
Note also that the gravitational potential can be expressed in 
terms of the escape speed $V_{esc}$ at the Alfv\'en radius $r_\star$, 
\[
{\cal V}  =-\frac{\nu ^{2} V_{\star}^{2}}{2R}\,, \qquad 
\nu=\frac{V_{esc}}{V_{\star}}\,,\qquad 
V_{esc}=\sqrt{\frac{2 {\cal GM} }{r_{\star}}}\,.
\]
Instead of using the three free functions of $\alpha$, (${\cal A}\,, 
\Psi_A$\,, $\Omega$), we found it more convenient to work instead with the 
three dimensionless functions of $\alpha$, ($g_1$\,, $g_2$\,, $g_3$),  
\begin{equation}\label{g1}
g_1\left( \alpha \right)= \int {\cal A}^{'2} d\alpha\,,
\end{equation}
\begin{equation}\label{g2}
g_2 \left( \alpha \right)=\frac{r_{\star}^2}{B_{\star}^2} \int \Omega ^2 \Psi_A ^2 d \alpha
\end{equation}
\begin{equation}\label{g3}
g_3\left(\alpha \right)=\frac{\Psi_A ^2 }{4\pi \rho_{\star}}\,. 
\end{equation}
Also, we shall indicate by ${\Pi}$ the total pressure in units of 
the magnetic pressure at the Alfv\'en surface on the polar axis, 
${B_{\star}^2}/{8 \pi}=\rho_{\star} V_{\star}^2/{2}$, 
\[
\Pi = \frac{8 \pi}{B_{\star}^2}  \left(
P+\frac{B^2}{8 \pi}\right)\,,
\]
such that,
\begin{equation}\label{pre1}
P=\frac{B_{\star}^2}{8 \pi} \left( {\Pi} + f_1 g_{1}^{'} + f_2 \alpha g_{1}^{'}+
f_3 \alpha g_{2}^{'} \right) \,.
\end{equation}
The functions $f_i(R)$, $i=1,2,3$ are given in Appendix A while 
all starred quantities refer to their respective values at the 
polar Alfv\'en point ($R=1$\,,$\alpha=0$). Hence, 

\[
{\cal A}^{'}\left(\alpha=0\right)=1\,,\; 
\Psi_A \left(\alpha=0\right)=\sqrt{4\pi \rho_{\star}}
\,,
\]
or,
\begin{equation}\label{kan}
g_1^{'}\left(\alpha=0\right)=1\,,\; g_3\left(\alpha=0\right)=1
\,.
\end{equation}

With assumptions (i)-(ii) and in this notation, the 
$\hat {r}-$ and $\hat{\theta}-$ components of the momentum equation become,
\begin{eqnarray}\label{rcomp}
\frac{\partial {\Pi}\left(R,\theta\right)}{\partial R}=f_6 g_{1}^{'} +\left(f_7 +\frac{F}{R} f_4\right)\alpha g_{1}^{'}+
\nonumber \\
\left(f_8+\frac{F}{R} f_5\right)\alpha g_{2}^{'}+f_9 g_3 \,,
\end{eqnarray}
\begin{equation}\label{thetacomp}
\frac{\partial {\Pi}\left(R,\theta\right)}{\partial \theta}=
2\cot{\theta}\left(f_4 \alpha g_{1}^{'} +f_5 \alpha g_{2}^{'} \right) 
\,.
\end{equation}
Next, by using $\alpha$ instead of $\theta$ as an independent variable, 
we may transform from 
pair of the independent variables ($R\,, \theta$) to pair of the 
independent variables ($R\,, \alpha$). With the following elementary 
relations valid for any differentiable function $\Phi$,
\begin{equation}\label{el1}
\frac{\partial {{\Phi}}\left(R,\theta\right)}{\partial R}=
\frac{\partial {{\Phi}}\left(R,\alpha\right)}{\partial R}
+\alpha \frac{F}{R} \frac{\partial {{\Phi}}\left(R,\alpha\right)}{\partial 
\alpha} \,,
\end{equation}
\begin{equation}\label{el2}
\frac{\partial {{\Phi}}\left(R,\theta\right)}{\partial \theta}=
2 \alpha \cot \theta 
\frac{\partial {{\Phi}}\left(R,\alpha\right)}{\partial \alpha} \,,
\end{equation}
we may transform Eqs. ({\ref{rcomp}), (\ref{thetacomp}}) 
into the following two equations:
\begin{equation}\label{1comp}
\frac{\partial {\Pi}\left(\alpha,R\right)}{\partial \alpha}=f_4 g_{1}^{'}+f_5 g_{2}^{'}
\,,
\end{equation}
\begin{equation}\label{2comp}
\frac{\partial {\Pi}\left(\alpha,R\right)}{\partial R}=
f_6 g_{1}^{'} +f_7 \alpha g_{1}^{'} +f_8 \alpha g_{2}^{'} +f_9 g_3 \,.
\end{equation}
By integrating Eq. ({\ref{1comp}}) we get  
${\Pi}=f_4 g_{1}+f_5 g_{2}+f_0$ where $f_0$ is an arbitrary function 
of $R$. From Eq. ({\ref{pre1}}) the pressure is  
\begin{equation}\label{pre}
P=\frac{B_{\star}^2}{8 \pi}\left(f_4 g_{1}+f_5 g_{2}+f_0+ 
f_1 g_{1}^{'} + f_2 \alpha g_{1}^{'}+
f_3 \alpha g_{2}^{'} \right) \,,
\end{equation}
or, 
\[
P=\frac{B_{\star}^2}{8 \pi} {\bf{Y}}  {\bf{P}}^{\dag} 
\,,
\] 
where {\bf P} and {\bf Y} are the (1 $\times$ 7) matrices, 
\begin{equation}\label{prenew}
{\bf{P}}=\left[ \; f_0\, \; f_4\, \; f_1\, \; f_2\, \; f_5\, \; f_3\, \; 
0 \, \right]\,,
\end{equation}
and 
\begin{eqnarray}\label{Yinew}
\begin{array}{l}
{\bf{Y}}=\left[\;Y_1\,\;Y_2\,\;Y_3\,\;Y_4\,\;Y_5\,\;Y_6\,\;Y_7\;\right]=
\\ \\
\left[\; 1\;g_1\,\;g_1^{'}\,\;\alpha g_{1}^{'}\,\;g_2\,\;\alpha g_{2}^{'}\,
\;g_3\,\right]
\,.
\end{array}
\end{eqnarray}
Substituting for ${\Pi}$ in  Eq. ({\ref{2comp}}) it follows, 
\begin{equation}\label{athr}
-f_9g_3-f_8\alpha g_{2}^{'}+f_5^{'}g_{2}-f_7\alpha g_{1}^{'}-
f_6g_{1}^{'} +f_4^{'}g_{1}+f_0^{'}=0
\,,
\end{equation}
an expression of the form \\
$X_7\left(R\right)Y_7\left(\alpha\right)+X_{6}\left(R\right)
Y_{6}\left(\alpha\right)+\cdots +X_{1}\left(R\right)
Y_1\left(\alpha\right)=0 
\,,$ 
\begin{eqnarray}\label{7-1}
\mbox{or} \,, 
{\bf{Y}} {\bf{X}}^{\dag} = {\bf{0}} 
\end{eqnarray}
with {\bf X} the (1 $\times$ 7) matrix
\begin{eqnarray}\label{Xinew}
{\bf{X}}=\left[\;X_1\,\;X_2\,\;X_3\,\;X_4\,\;X_5\,\;X_6\,\;X_7\;\right]=
\nonumber\\
\left[\;f_0^{'}\,\;f_4^{'}\,\;-f_6\,\;-f_7\,\;f_5^{'}\,\;-f_8\,\;-f_9\; \right] 
\,.
\end{eqnarray}

\subsection{Systematic construction of classes of meridionally selfsimilar
MHD outflows}

\begin{table*}
\centering
\begin{minipage}{140mm}
\caption{Meridionally Selfsimilar Models}
\begin{tabular}{@{}cllll@{}}
Case& $g_1(\alpha)$ & $g_2(\alpha)$ & $g_3(\alpha)$ & constants\\
\hline
(1) & $\alpha$ 
& $\lambda^2 \alpha$ 
& $1+\delta \alpha$ &\\
(2) & $\alpha$
& $\xi \alpha+{\mu \alpha^{\epsilon}/\epsilon}$ 
& $1+\delta \alpha +\mu \delta_{0} \alpha^{\epsilon}$ &
$\epsilon \neq 0,1 ,\mu \neq 0 $
\\
 (3) & $\alpha$  & $\xi \alpha + \mu \alpha \ln {\alpha} $  
 & $1+\delta \alpha +\mu \delta_{0} \alpha \ln {\alpha}$ &
$\mu \neq 0 $ \\
(4) & $\alpha_0 e^{{\alpha\over \alpha_0}}$ 
& $\lambda e^{{\alpha\over \alpha_0}}$ 
& $1+\delta \alpha e^{{\alpha\over \alpha_0}}+
\mu \left(e^{{\alpha\over \alpha_0}}-1\right)$ & \\
(5) & $\frac{\alpha_0}{\epsilon} \mid\frac{\alpha}{\alpha_0}-1 
\mid ^{\epsilon-1} 
\left(\frac{\alpha}{\alpha_0}-1\right)$ 
& $\xi\mid\frac{\alpha}{\alpha_0}-1 \mid ^{\epsilon}$ 
& $ 1+\delta \mid\frac{\alpha}{\alpha_0}-1 \mid ^{\epsilon}+\mu 
\mid\frac{\alpha}{\alpha_0}-1 \mid ^{\epsilon-1}-\delta -\mu$ &
$\epsilon \neq 0,1 $\\
(6) & $-\alpha_0 \ln{\mid \frac{\alpha}{\alpha_0}-1\mid}$ 
& $\xi \ln{\mid \frac{\alpha}{\alpha_0}-1\mid}$ 
& $1+\delta \ln{\mid \frac{\alpha}{\alpha_0}-1\mid}+
\mu \frac{\alpha}{\alpha_0 \left(\alpha -\alpha_0\right)}$ & \\
(7) & $\frac{\alpha}{1-\alpha_0}$
& $\mu \ln{\frac{\alpha}{\alpha_0}}+ \xi \alpha$
& $1+\delta \left(\alpha-\alpha_0 \right) +
\mu \delta_{0} \ln{\frac{\alpha}{\alpha_0}}$ &
$\mu \neq 0 $ \\
(8) & $\frac{\alpha_0}{\epsilon 
\left(1-\alpha_0\right)}\left(\frac{\alpha}{\alpha_0}
\right)^{\epsilon}$
& $\lambda_1 \alpha^{\epsilon}+\lambda_2 \alpha^{\epsilon-1}$ 
& $1+\delta_1\left( \alpha^{\epsilon}-\alpha_0^{\epsilon}\right) 
+\delta_2\left(\alpha^{\epsilon-1}-\alpha_0^{\epsilon-1}\right)$ &
$\epsilon \neq 0,1 $\\
(9) & $\frac{\alpha_0}{1-\alpha_0}\ln{\frac{\alpha}{\alpha_0}}$
& $\lambda_1 \ln{\frac{\alpha}{\alpha_0}}+{\lambda_2 \over \alpha}$
& $1+\delta_1 \ln{\frac{\alpha}{\alpha_0}} +
\delta_2\left(\frac{1}{\alpha}-\frac{1}{\alpha_0}\right)$ &\\
\hline
\end{tabular}
\end{minipage}
\end{table*}
In (Vlahakis \& Tsinganos 1997) the following simple theorem was proved:\\
{\it Theorem}: If $F_n(\alpha)$, $Y_{i}(\alpha)$, $X_{i}(R)$, $i=1,2,\cdots ,n$ are 
arbitrary functions of the independent variables $\alpha$ and $R$ and 
\begin{equation}\label{1}
F_n\left(\alpha\right)=Y_{1}\left(\alpha\right)X_{1}\left(R\right)+
\cdots +Y_{n}\left(\alpha\right)X_{n}\left(R\right),
\end{equation}
then, there exist constants \( c_{1},c_{2},\ldots , c_{n} \) such that,
\begin{equation}\label{2}
F_n\left(\alpha\right)=c_{1}Y_{1}\left(\alpha\right)+c_{2}Y_{2}\left(\alpha\right)+ \cdots 
+c_{n}Y_{n}\left(\alpha\right). 
\end{equation}

Consider then a relation of the form,
\begin{equation}
\label{n}
X_n\left(R\right)Y_n\left(\alpha\right)
+\cdots +X_{1}\left(R\right)Y_1\left(\alpha\right)=0 \,.
\end{equation}
Regarding the first term of the sum there are evidently only two 
possibilities. Either, 
\begin{enumerate}
\item
$X_n\left(R\right)=0$ for every $R$, in which case (indicated by 
the digit "0") we have 
\begin{eqnarray}
\qquad\quad X_{n-1}\left(R\right)Y_{n-1}\left(\alpha\right)+\cdots +
X_{1}\left(R\right)Y_1\left(\alpha\right)=0
\nonumber
\,,
\end{eqnarray}
or, 
\item $X_n\left(R\right) \neq 0$, in which case (indicated by the 
digit "1") we have 
\begin{eqnarray}
\qquad\quad Y_n\left(\alpha\right)=-\frac{X_1\left(R\right)}{X_n\left(R\right)}Y_1
\left(\alpha\right)-\cdots
-\frac{X_{n-1}\left(R\right)}{X_n\left(R\right)}Y_{n-1}\left(\alpha\right)
\,.\nonumber
\end{eqnarray}
\begin{itemize}
\item[] Then, according to the theorem stated in the beginning of this 
section, there are constants 
$\mu_i^{\left(n\right)},i=1,2,\cdots,n-1$ such that
$Y_n\left(\alpha\right)=\sum_{i=1}^{n-1} \mu_i^{\left(n\right)} 
Y_i\left(\alpha\right)$. This gives a condition between the functions of 
$\alpha$. Substituting this in the initial sum we find:
\end{itemize}
\begin{eqnarray}
\begin{array}{l}
 \left[ X_{n-1}\left(R\right)+
\mu_{n-1}^{\left(n\right)}X_n\left(R\right)\right]
Y_{n-1}\left(\alpha\right)+
\\ \\
\left[X_{n-2}\left(R\right)+\mu_{n-2}^{\left(n\right)}X_n\left(R\right)\right]
Y_{n-2}\left(\alpha\right)+\cdots \\ \\
+ \left[X_1\left(R\right)+\mu_1^{\left(n\right)}X_n\left(R\right)\right]
Y_1\left(\alpha\right)=0
\,.
\end{array}
\end{eqnarray}
\end{enumerate}
Hence, in both cases we find a sum with $n-1$ terms.
Following this algorithm at the end we 'll have only one term.
Since for each product we have the above two possibilities, totally we obtain 
$2^n$ cases.
Each of them corresponds to a set "xx$\cdots$xx" with $x=1, 0\,$($n$ digits).
The number of "1" digits is the number of conditions between functions of 
$\alpha$ while the number of "0" digits is the number of conditions between 
functions of $R$.\\

Now, following this method from Eq. ({\ref{athr}}) 
we get $2^7$ solutions. Each of them corresponds to 
a set "xxxxxxx" with $x$ either 1, or, 0. Of those numbers:
\begin{enumerate}
\item The first digit is always "1" (because $X_7 \neq 0$).
\item The last digit is always "0" (because $Y_1 \neq 0$).
\item Since ${\cal A'}\neq 0$, it follows that $g_1'\neq 0$ and thus $g_1$ 
cannot be a constant. Hence, the function $Y_2=g_{1}$ cannot be 
proportional to $Y_1$ and therefore all numbers always have "00" at 
the end.
\item We have totally six unknown functions: the three functions of 
$R$, ($G,M,f_0$) and the three functions of $\alpha$, 
($g_{1}, g_{2}, g_{3}$). 
On the other hand, the number of conditions between the functions of $R$ 
(their number is equal to the number of digits "0") and the functions of 
$\alpha$ (their number equals to the number of digits "1") 
in each one of the sets "xxxxxxx" is seven. 
It follows that the system of ($G,M,f_0$) and 
($g_{1}, \,g_{2},\,g_{3}$) is overdetermined.
Note however that since the forms of the functions $X_i(R)$ are much more 
complicated than the forms of the functions $Y_i(\alpha )$,  we choose sets 
"xxxxxxx" with at most three "0's" because in the case of 4 or more "0's" 
we have correspondingly 4 or more relations between the 3 functions of $R$, 
which in general overdetermines the system of ($G,M,f_0$). We then shift the 
problem of overdetermination of the problem to the set of the 3 functions of 
$\alpha$, ($g_{1},\,g_{2},\,g_{3}$) which need to satisfy 4 relations.
In this system however, it is possible to choose the constants $\mu_i^{\left(j\right)}$ 
such that a consistent solution for the functions of $\alpha$ can be finally 
constructed.  
\end{enumerate}
Altogether, then and with these considerations in mind, 
from the $2^7=128$ possible cases we end up with only five: 
1011100, 1101100, 1110100, 1111000, 1111100. 
For each of one of those sets we can solve the system for 
$g_{1}\,,g_{2}\,,g_{3}$, as it is shown in the example of the next 
section.\\ 
From a different perspective, $g_{1}(\alpha),\,g_{2}(\alpha),\,
g_{3}(\alpha)$ are vectors in a 3D $\alpha$-space with basis  
vectors [$ u_1(\alpha)\,,  u_2(\alpha)\,,  u_3(\alpha)$]. 
This space contains all vectors $g_i(\alpha)$, $i=1,2,3$ subject to 
the $\theta$-selfsimilarity constraint manifested by Eq. (\ref{athr}), i.e.,  
that for a given such set $g_i(\alpha)$, $i=1,2,3$, the vectors 
$1,\alpha g_{1}^{'}(\alpha)\,, \alpha g_{2}^{'}(\alpha)$ and 
$g_{1}^{'}(\alpha)$ also belong to the same space. 
Each of the resulting functions $g_i(\alpha)$, $i=1,2,3$ are 
then a linear combination of the 
basis vectors $ u_1(\alpha),  u_2(\alpha),  u_3(\alpha)$.
In the following, we choose $ u_1=1$, $ u_2= g_1(\alpha )$. 
All such sets of basis vectors give all possible meridionally 
selfsimilar solutions. Therefore, collecting all possibilities, we 
end up with the classes of solutions shown in Table 1.
Note that in the last three cases $A^{'}\left(\alpha=0\right)\neq 1$, 
but one can say that the starred quantities refer to values at the point 
$R=1,\alpha=\alpha_0<1$.

In all nine cases of Table 1, from Eqs. ({\ref{g1}}), ({\ref{g2}}), 
({\ref{g3}}) we may find easily the forms of the free integrals from 
the relations, 
\\
\begin{equation}\nonumber
A=\frac{B_{\star}r_{\star}^{2}}{2} \int _0 ^{\alpha} \sqrt{g_1^{'}} d\alpha,
\;\;\Psi_{A}={\sqrt{4\pi\rho_{\star} g_3 }} 
\,,
\end{equation}
\begin{equation}{}
\Omega=\frac{V_{\star}}{r_{\star}} \sqrt{\frac{g_2^{'}}{g_3}}\,,\;\; 
L=r_{\star} V_{\star} \alpha \sqrt{\frac{g_2^{'}}{g_3}}
\,,
\end{equation}
while by substituting $g_1, g_2, g_3$ in Eqs. ({\ref{pre}}), 
({\ref{athr}}),  the corresponding \underline{ordinary} differential 
equations for the jet radius $G(R)$, Alfv\'en number $M(R)$ and 
pressure component $f_0(R)$ are found from the $R$-relations, as it 
is illustrated in the following section.
\\
From the perspective of the $\alpha$-space, in each one of the cases of 
Table 1 there exists a $3 \times 7$ matrix ${\bf {K}}$ such that 
\begin{equation}{}
{\bf{Y}}=\left[\;{ u_1}\,\;{ u_2}\,\;{u_3}\;\right] 
{\bf{K}}
\,,
\end{equation}
so that from  Eq. ({\ref{7-1}}),
\[
\left[\;{ u_1}\,\;{u_2}\,\;{ u_3}\;
\right]\,{\bf {K}} {\bf{X}}^{\dag}={\bf{0}} \,.
\] 
If ${u_i}$ are linearly independent then 
\[
{\bf{K}} {\bf{X}}^{\dag}={\bf{0}} \,.
\]
These three equations are the \underline{ordinary} differential equations for the 
functions of $R$ in each model while the pressure is,
\[
P=\frac{B_{\star}^2}{8 \pi} \left[\;{ u_1}\,\;{ u_2}\,\;{ u_3}
\;\right] {\bf{K}} {\bf{P}} ^{\dag} 
= \frac{B_{\star}^2}{8 \pi} \left( P_0 + P_1g_1+P_2u_3 \right)
\,, \] 
where
\[
{\bf K} {\bf P}^{\dag}= [P_0 \; P_1 \; P_2]^{\dag} 
\,.
\]
The first two cases of Table 1 are of some interest. The first,
is a degenerate one with $ u_3 = 0$ and 
the following form of the free integrals:
\begin{eqnarray}\label{TTS}
\begin{array}{l}
A=\frac{B_{\star}r_{\star}^{2}}{2}\alpha \,, \;
\Psi_{A}=\sqrt{4\pi\rho_{\star}\left(1+\delta \alpha\right)} \,,
\\ \\
\Omega=\frac{\lambda V_{\star}}{r_{\star}}\frac{1}{\sqrt{1+\delta 
\alpha}}
\,.
\end{array}
\end{eqnarray}
\begin{table*}
\centering
\begin{minipage}{140mm}
\caption{Meridionally Selfsimilar Radial Models }
\begin{tabular}{@{}clll@{}}
Case& $g_1(\alpha)$ & $g_2(\alpha)$ & $g_3(\alpha)$ \\
\hline
(1) & $ -\ln{\mid 1-\alpha \mid }$ & $0$ & $1$ \\
(2) & $ \mu \int_{}^{} \frac{\alpha^{\epsilon}}{1-\alpha}d\alpha-\ln{\mid 1-\alpha \mid }$
& $\lambda^2 {\alpha^{\epsilon}\over \epsilon}$ & $1+\delta \alpha^{\epsilon}$ \\
(3) & $\mu_1 \ln{\mid 1-\alpha \mid }+\mu_2 \int \frac{\ln{\alpha }}{1-\alpha} d\alpha $ &
$\lambda \ln{\alpha }$ & $\delta_1+\delta_2 \ln{\alpha }$ \\
(4) & $g_1(\alpha)\neq \mu_1 \ln{\mid 1-\alpha \mid }+\mu_2$ &$0$ & $\delta g_{1}^{'} \left(1-\alpha \right)+\lambda $ \\
(5) & $\mu \ln{\mid 1-\alpha \mid }$ & $g_{2}(\alpha)\neq (\mu, \mu_1 
\ln{\alpha}+\mu_2,\mu_1 \alpha^{\mu_2}+\mu_3) $ & $\delta$ \\
\hline
\end{tabular}
\end{minipage}
\end{table*}
This is a special case of the more general following case (2) for $\mu =0$ 
(and $\xi =\lambda^2$) and has been studied in detail in ST94 
and Trussoni et al (1997). 
It is the single case where we have only two 
conditions between the functions of $R$, so that the third relation 
between the unknown functions $G, \, M, \,f_0$ is freely chosen. In 
Trussoni et al (1997) this corresponds to an {\it a priori} 
specification of the shape of the poloidal streamlines, while in ST94 
in an {\it a priori} imposed relationship between 
the spherically and nonspherically symmetric components 
of the pressure. This last case leads to a generalized polytropic-type 
relation between pressure and density of the form,
\begin{equation}\label{Berng}
\frac{P(\alpha,R)}{P(0,R)} = \mbox{ function of  } \;
\frac{\rho(\alpha,R)}{\rho(0,R)}
\,.
\end{equation}
As a result, a Bernoulli-type constant exists and, among others, this 
constant gives a quantitative criterion for the transition of an 
asymptotically conical wind from an inefficient magnetic rotator to an 
asymptotically cylindrical jet from an efficient magnetic rotator.\\
The second case 
with $\epsilon \neq 0,1 ,\mu \neq 0 $ has $u_2 = \alpha$, 
$u_3 = \alpha^{\epsilon}$. The corresponding form of the 
free integrals is : 
\begin{equation}\label{V}
\begin{array}{l}
A=\frac{B_{\star}r_{\star}^{2}}{2}\alpha, \; 
\Psi_{A}=\sqrt{4\pi\rho_{\star}\left(1+\delta \alpha+\mu \delta_{0} 
\alpha^{\epsilon}\right)} 
\,, \\ \\
\Omega=\frac{V_{\star}}{r_{\star}}\sqrt{\frac{\mu \alpha^{\epsilon-1}+\xi}{
1+\delta \alpha+\mu \delta_{0} \alpha^{\epsilon}}}\,.
\end{array}
\end{equation}
This is a new case which emerged from the present systematic 
construction. The corresponding differential equations 
are derived in detail in the example of the next section 
where the solution is briefly analysed. 
\\
In the special configuration with $G=R \Leftrightarrow \alpha=
\sin^2 \theta$, the field and stream lines on the poloidal plane 
are radial and we find five cases shown in Table 2.\\
The first case is a {\it degenerate} one, wherein there is only one 
condition between the unknown functions $M(R)\,,f_0(R)$. 
Thus, a second relation between $M(R)-f_0(R)$ can be imposed 
{\it a priori}, for example, a polytropic relation between pressure 
and density. This last possibility leads precisely to Parker's (1963) 
classical solar wind solution with a radial and nonrotating outflow. 
All other cases (2)-(5) are {\it non-degenerate}, i.e., there are two 
relations between $M(R)-f_0(R)$.\\
The second case has been analysed in detail in Lima et al (1996) and 
corresponds to a radial but heliolatitudinally depended outflow. 
In addition $\mu=-1$, $\epsilon =1$ this case coincides with 
(1) in Table 1 for radial poloidal streamlines. Note 
that a common feature of all rotating cases with radial stream lines 
on the poloidal plane is that they cannot be extended in all the 
poloidal plane, for sufficiently fast magnetic rotators. 
For example, in the model of Lima et al. (1996) the pressure 
becomes negative at some colatitude $\theta_{max}$, for large values 
of rotation. This is basically due to the fact that with the poloidal 
magnetic field dropping like $1/R^{2}$ and the azimuthal field 
dropping like $1/R$, the magnetic pressure drops like $1/R^4$ and 
by itself alone cannot balance the magnetic tension which 
drops like $1/R^3$; a strong pressure gradient is then needed from the 
pole towards the equator to balance the magnetic pinching. In fast 
magnetic rotators this pressure  gradient is so strong that it leads 
to negative values of the pressure at angles $\theta > \theta_{max}$. 
A collimated outflow with uniform asymptotic conditions is the only 
way left for an everywhere valid outflow from an efficient magnetic rotator 
(Heyvaerts \& Norman 1989, ST94).  
\subsection{Example of a new model for a meridionally self-similar MHD outflow} 

Let us illustrate the previous construction with the example 1101100 
obtained from the present case with $n=7$. This number means the 
following:\\ 
Since the first digit is 1, there are six constants 
$\mu_i^{\left(7\right)},i=1,2,\cdots,6$ such that the following relation 
holds between the functions $Y_i(\alpha)$, i=1,2, ..7, 
\begin{eqnarray} 
\label{Y7}
Y_7=\sum_{i=1}^{6} \mu_i^{\left(7\right)}Y_i \,,
\qquad \mbox{($\alpha$-relation-1)} 
\,.
\end{eqnarray}
Substituting this expression of $Y_7$ in the initial relation 
Eq. (\ref{7-1}) 
between the functions ($X_i$,  $Y_i$), i=1,..7, we obtain 
\begin{eqnarray}
\begin{array}{l}
\label{XY6}
\left(X_6+\mu_6^{\left(7\right)}X_7\right)Y_6+
\left(X_5+\mu_5^{\left(7\right)}X_7\right)Y_5+\cdots+
\\ \\ 
\left(X_1+\mu_1^{\left(7\right)}X_7\right)Y_1=0 \,.
\end{array} 
\end{eqnarray} 
Now the second digit is again 1 and thus there are five constants 
$\mu_i^{\left(6\right)},i=1,2,\cdots, 5$ such that
\begin{eqnarray}
\label{Y6}
Y_6=\sum_{i=1}^5 \mu_i^{\left(6\right)}Y_i\,, 
\qquad \mbox{($\alpha$-relation-2)} 
\,,
\end{eqnarray}
while substituting this relation in Eq. (\ref{XY6}) we obtain, 
\begin{eqnarray}
\begin{array}{l}
\label{XY5}
\left[\left(X_5+\mu_5^{\left(7\right)}X_7\right)+
\mu_5^{\left(6\right)}\left(X_6+\mu_6^{\left(7\right)}X_7\right)\right]Y_5+
\cdots \\ \\
+\left[\left(X_1+\mu_5^{\left(7\right)}X_7\right)+
\mu_1^{\left(6\right)}\left(X_6+\mu_6^{\left(7\right)}X_7\right)\right]Y_1
=0 \,.
\end{array}
\end{eqnarray}
The third digit is 0 and hence 
\begin{eqnarray}
\begin{array}{l}
\label{X1}
\left(X_5+\mu_5^{\left(7\right)}X_7\right)+
\mu_5^{\left(6\right)}\left(X_6+\mu_6^
{\left(7\right)}X_7\right)=0 
\\ \\
\mbox{($R$-relation-1)} 
\end{array}
\end{eqnarray}
a relation between the functions of $R$. With the help of Eq. (\ref{X1}), 
Eq. (\ref{XY5}) now reduces to,
\begin{eqnarray}
\label{XY4}
\sum_{i=1}^4 
\left[\left(X_i+\mu_i^{\left(7\right)}X_7\right)+
\mu_i^{\left(6\right)}\left(X_6+\mu_6^{\left(7\right)}X_7\right)\right]Y_i=0 \,.
\end{eqnarray}
The fourth digit is 1 and thus there are three constants 
$\mu_i^{\left(4\right)},i=1,2,3$ such that
\begin{eqnarray}
\label{Y4}
Y_4=\sum_{i=1}^3  \mu_i^{\left(4\right)}Y_i\,,
\qquad \mbox{($\alpha$-relation-3)} 
\,.
\end{eqnarray}  
Substituting this relation in Eq. (\ref{XY4}) we obtain 
\begin{eqnarray}
\begin{array}{l}
\label{XY3}
\sum_{i=1}^3 \left\{ \left[\left(X_i+\mu_i^{\left(7\right)}X_7\right)+
\mu_i^{\left(6\right)}\left(X_6+ 
\mu_6^{\left( 7 \right) }X_7\right)\right] + \right. 
\\ \\
\left. 
\mu_i^{\left(4\right)}\left[\left(X_4+\mu_4^{\left(7\right)}X_7\right)+
\mu_4^{\left(6\right)}\left(X_6+\mu_6^
{\left(7\right)}X_7\right)\right] \right\}Y_i=0 \,.
\end{array}
\end{eqnarray}
The fifth digit is 1 and there are two constants $\mu_i^{\left(3\right)},i=1,2$ 
such that
\begin{eqnarray}
\label{Y3}
Y_3=\mu_1^{\left(3\right)}Y_1+\mu_2^{\left(3\right)}Y_2 \,,
\qquad \mbox{($\alpha$-relation-4)} 
\,.
\end{eqnarray}
Substituting this in Eq. (\ref{XY3}) we find a relation involving 
$Y_1$ and $Y_2$. 
Finally, we must put equal to zero the multipliers of 
$Y_1, Y_2$ in this relation because the two remaining digits are 0.
So we have 
\begin{eqnarray}
\begin{array}{l}
\label{X23}
\left[\left(X_i+\mu_i^{\left(7\right)}X_7\right)+
\mu_i^{\left(6\right)}\left(X_6+ 
\mu_6^{\left( 7 \right) }X_7\right)\right] + 
\\ \\
\mu_i^{\left(4\right)}\left[\left(X_4+\mu_4^{\left(7\right)}X_7\right)+
\mu_4^{\left(6\right)}\left(X_6+\mu_6^
{\left(7\right)}X_7\right)\right]+  
\\ \\
\mu_i^{\left(3\right)} \left(
\left[\left(X_3+\mu_3^{\left(7\right)}X_7\right)+
\mu_3^{\left(6\right)}\left(X_6+ 
\mu_6^{\left( 7 \right) }X_7\right)\right] + \right. 
\\ \\
\left.
\mu_3^{\left(4\right)}\left[\left(X_4+\mu_4^{\left(7\right)}X_7\right)+
\mu_4^{\left(6\right)}\left(X_6+\mu_6^
{\left(7\right)}X_7\right)\right] \right)
=0\,,
\\ \\
\mbox{ for } i=1,2 \mbox{($R$-relations-2,3)} \,.
\end{array}
\end{eqnarray}
These last two equations together with Eq. (\ref{X1}) are the three 
equations for the functions of $R$. On the other hand, Eq. (\ref{Y7}), 
Eq. (\ref{Y6}), Eq. (\ref{Y4}) and Eq. (\ref{Y3}) are  
four relations among the three functions of $\alpha$. These relations 
of the functions of $\alpha$ 
[Eqs. ({\ref{Y3}}), ({\ref{Y4}}), ({\ref{Y6}}), ({\ref{Y7}})]
are equivalent to the system:
\begin{eqnarray}
\nonumber
\left.
\begin{array}{llll}
Y_3=c_1 Y_1 + c_2 Y_2 \\ Y_4=c_3 Y_1 + c_4 Y_2 \\ Y_6=c_5 Y_1 + c_6 Y_2 + c_7 Y_5 \\
Y_7=c_8 Y_1 + c_9 Y_2 +c_{10} Y_5 
\end{array}
\right\} \Leftrightarrow \left\{
\begin{array}{llll}
g_1^{'}=c_1 + c_2 g_1 \\ \alpha g_1^{'}=c_3 + c_4 g_1 \\
\alpha g_2^{'}=c_5 + c_6 g_1 +c_7 g_2 \\ g_3 =c_8  + c_9 g_1 +c_{10} g_2
\end{array}
\right.
\end{eqnarray}
Note that we renamed the constants and also used Eq. ({\ref{Yinew}}). 
From the first, if $c_2 \neq 0$ it follows that 
$g_1=-{c_1}/{c_2}+c e^{c_2 \alpha}$. Then, from the second $c=0$ and 
hence $g_1=-{c_1}/{c_2}$.
But $g_1$ cannot be a constant. Thus, $c_2 =0$ while the first two 
equations combined with Eq. ({\ref{kan}}) give   
$g_1=\alpha + c_{11}$ while the third has the solutions:\\ \\
$  
g_2= \left\{\begin{array}{lll}
\frac{c_6}{1-c_7}\alpha + c_{12} + c_{13} \alpha^{c_7}\,, \; 
\mbox{if} \; \; c_7 \neq 0,1 \\
c_6 \alpha \ln{\alpha} +c_{14} +c_{15} \alpha  \,, \; \mbox{if} \;\; 
c_7 =1 \\
c_6 \alpha +c_{16} \ln{\alpha} +c_{17} \,, \; \mbox{if} \;\; c_7 =0\,.
\end{array} \right. $
\\ \\
For the first possibility, we have finally the second case of Table 1 :\\
\\
$
\begin{array}{lll}\label{g123}
g_1=\alpha \\g_{2}=\xi \alpha+{\mu \alpha^{\epsilon}\over \epsilon}\;, \\
g_{3}=1+\delta \alpha +\mu \delta_{0} \alpha^{\epsilon}
\end{array} \epsilon \neq 0,1 
$
\\
\\
where we have absorbed the constants $c_{11}, c_{12}$ in the 
unknown function $f_0$, 
$c_{11} f_4+ c_{12} f_5 +f_0 \rightarrow f_0$, 
Eqs. (\ref{pre}), (\ref{athr}).\\

After substituting these values of $g_1, g_2, g_3$ in 
Eqs. ({\ref{pre}}) - ({\ref{athr}}), we find that 
\begin{eqnarray}\label{feq}
\begin{array}{l}
[f_0^{'}-f_6-f_9]+
\left[f_4^{'}+\xi f_5^{'}-f_7-\xi f_8 -\delta f_9 \right]\alpha +
\\ \\
\mu \left[{f_5^{'}}/{\epsilon}-f_8-\delta_0 f_9 \right]
\alpha^{\epsilon}=0 
\,,
\end{array}
\end{eqnarray}
and
\begin{eqnarray}\label{peq}
\begin{array}{l}
P=\frac{B_{\star}^2}{8 \pi}\left(P_0+P_1 \alpha +P_2  
\alpha^{\epsilon}\right) 
\\ \\
= \frac{B_{\star}^2}{8 \pi} \left[f_0+f_1+\left(
f_4+\xi f_5 +f_2+\xi f_3 \right) \alpha + 
\mu \left(\frac{f_5}{\epsilon}+f_3\right) \alpha^{\epsilon} \right]
\,.
\end{array}
\end{eqnarray}
By setting equal to zero the three expressions in the square 
brackets of Eq. (\ref{feq}) (since $\mu \neq 0$ and 
$1\,, \alpha\,, \alpha^{\epsilon}$ are 
linearly independent vectors in the $\alpha$-space for $\; \epsilon \neq 0,1\;$)
we find the three R-relations for the functions $G(R)$, $M(R)$, $f_0(R)$ 
(which are the same with Eqs.({\ref{X1}}),({\ref{X23}})).
Using the functions $f_4,F$ and 
the definitions of $P_0$ 
and $P_1$ we obtain five, \underline{first order}, ordinary 
differential equations for $G(R)$, $F(R)$, $M(R)$ and the two pressure 
components $P_1(R)$ and $P_0(R)$,
\begin{equation}
\begin{array}{l}
{d G^2\over dR} = - {{F-2}\over R}G^2
\,,
\end{array}
\end{equation}

\begin{eqnarray}
\begin{array}{l}
\frac{dF}{dR}=\frac{F}{1-M^2}\frac{d M^2}{dR}-\frac{F\left(F-2
\right)}{2R}-
\\ \\
\frac{F^{2}-4}{2R \left(1-M^2 \right)}-
\frac{2G^2RP_{1}}{1-M^2}-
\\ \\
\frac{2\xi R}{M^2 \left(1-M^2 \right)^{3}}\left[\left(2M^2-1\right)
G^4-M^4+2M^2\left(1-G^2\right)\right]
\end{array}
\end{eqnarray}

\begin{eqnarray}
\begin{array}{l}
\frac{d M^2}{dR} = \frac{M^2 \left(1-M^2 \right)}{ \left(2M^2-1
\right) G^4-M^4}\left\{
-\epsilon \delta_0 \nu^{2} \right. \frac{G^2 \left(1-
M^2\right)}{R^{2}}+
\\ \\
\left.
\frac{F-2}{R}\left[\left(\epsilon+1\right)M^2-\left(\epsilon-1\right)
G^4\right]\right\}
\end{array}
\end{eqnarray}

\begin{eqnarray}
\begin{array}{l}
\frac{dP_{1}}{dR}=- \left[\frac{F^{2}-4}{2R^{2}G^2}+2\xi
\frac{\left(1-G^2\right)^
{2}}{G^2\left(1-M^2 \right)^{3}}\right] \frac{d M^2}{dR}-
\\ \\
\frac{M^2 F}{2R^{2} G^2}\frac{dF}{dR}-
\frac{\delta \nu^{2}}{R^{2}M^2}-
\frac{M^2 \left(F^{2}-4\right)\left(F-4\right)}{4R^{3}G^2}+
\\ \\
\xi\frac{\left(F-2\right)\left[\left(2M^2-1\right)
 G^4-M^4\right]}{RG^2M^2\left(1-M^2 \right)^{2}}
\end{array}
\end{eqnarray}

\begin{eqnarray}
\begin{array}{l}
\frac{dP_{0}}{dR}=-\frac{2}{G^4}\frac{d M^2}{dR}-\frac{\nu^{2}}{R^{2}M^2}-
\frac{2M^2\left(F-2\right)}{RG^4}
\end{array}
\end{eqnarray}
\begin{figure}
\centerline{\psfig{file=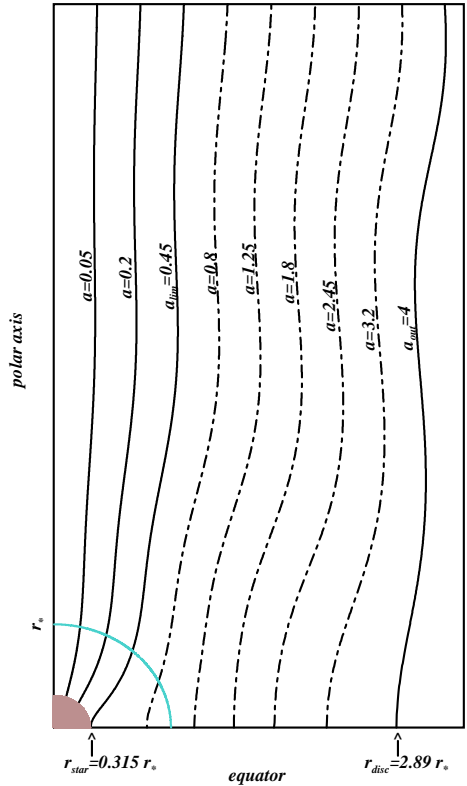,height=13.0truecm,angle=360}}
\caption{Poloidal field and streamlines close to the stellar base for 
the asymptotically cylindrical $\theta$-self similar model of case (2) 
from Table 1, for the following set of parameters: $\epsilon=0.5$,
$\nu^2=2{\cal G M}/ r_{\star} V_{\star}^2=10$,
$\delta \nu^2 =3.5$, $\delta_{0} \nu^2=0.1$, $\xi=-10$, $\mu=20$,
$p_{\star}=\left({dM^2}/{dR}\right)_\star=1.6$,$F_\star=1.1$.}
\end{figure}
\begin{figure}
\centerline{\psfig{file=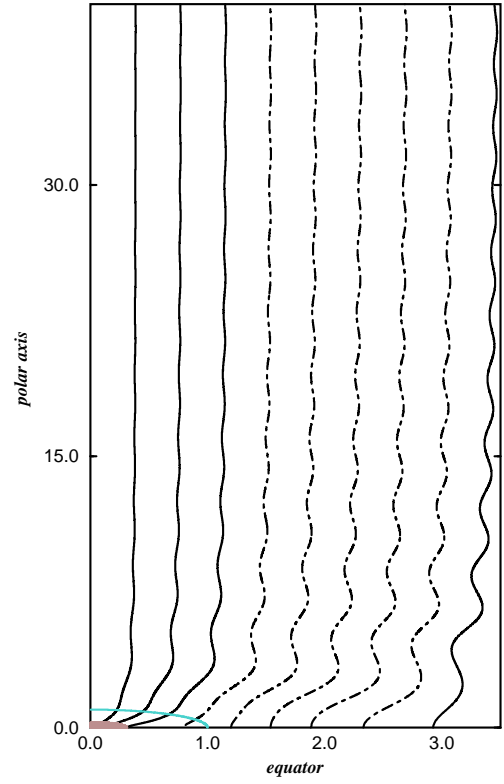,height=13.0truecm,angle=360}}
\caption{Poloidal field and streamlines as in Fig (1), but in an enlarged 
scale to show the asymptotical collimation reached after the oscillations 
have decayed.}
\end{figure}
Note that the third pressure component $P_2(R)$ is given explicitly in 
terms of $M$ and $G$ ($f_3$ and $f_5$). An integration of the above 
set of equations will give the complete solution. However, this 
exercise is rather complicated since any physically accepted solution 
should pass through the various MHD critical points (Tsinganos et 
al 1996). This undertaking, 
together with a discussion of the solution and application to 
collimated outflows is the subject of the next paper. 

It is worth mentioning at this point that our analysis of model (2) of 
Table 1 shows that mainly cylindrically collimated solutions are obtained. 
The set of Figures (1-2) illustrates such a typical 
solution for a representative set of the constants describing the particular model.
This solution crosses the Alfv\'en surface for appropriate values of the 
slope of the square of the Alfv\'en number $p_\star = 
\left( {dM^2}/{dR}\right)_\star$, the expansion function $F_\star$ 
and $P_{1\star}$ which satisfy the Alfv\'en regularity condition 
(Heyvaerts \& Norman 1989, ST94) which is easily obtained from 
Eq. (\ref{f4}) of Appendix A at ($R=G=M=1$), i.e., 
\begin{equation}\label{t-alfven}
F_{\star} p_{\star} = 2 f_{4\star}
\,.
\end{equation}
The nonspherically symmetric part of the pressure $P_{1\star}$ is obtained from its 
definition while the functions $f_{3\star}\,,f_{5\star}$ are calculated for $R=1$ using the 
L'Hospital rule.
Figs. (1,2,3) correspond to the set $F_{\star}=1.1$ and $p_\star=1.6$. 
Note that after the Alfv\'en star-type critical point is crossed, the modified by 
self-similarity X-type fast critical point (Tsinganos et al 1996) may be 
crossed by further adjusting appropriately the triplet of the variables 
$(F_\star\,, p_\star\,, P_{1\star})$.  
It suffices to note that solutions crossing only the Alfv\'en surface do not 
differ qualitatively from those which in addition cross the modified by the 
present meridional selfsimilarity fast critical surface. 

Fig. (1) shows the shape of the streamlines on the poloidal plane and 
close to the Alfv\'en surface. The cylindrical asymptotical shape of the 
poloidal streamlines is shown in the enlarged scale of Fig (2). Note also 
the constant wavelength but the decaying with distance amplitude of the 
oscillations, in full agreement with the analysis in Vlahakis \& Tsinganos 
(1997). At the last shown fieldline $\alpha_{out}=4$, the toroidal fields vanish 
$B_{\phi}=0$, $V_{\phi}=0$.For $\alpha>\alpha_{out}\,,\Omega^2$ becomes negative ,
so there is no solution there.
The same oscillatory behaviour can be seen in the fieldlines which are not 
rooted on the star but they are perpendicular to a thin disk around it 
(dotted curves in Figs 1, 2.) 
\begin{figure}
\centerline{\psfig{file=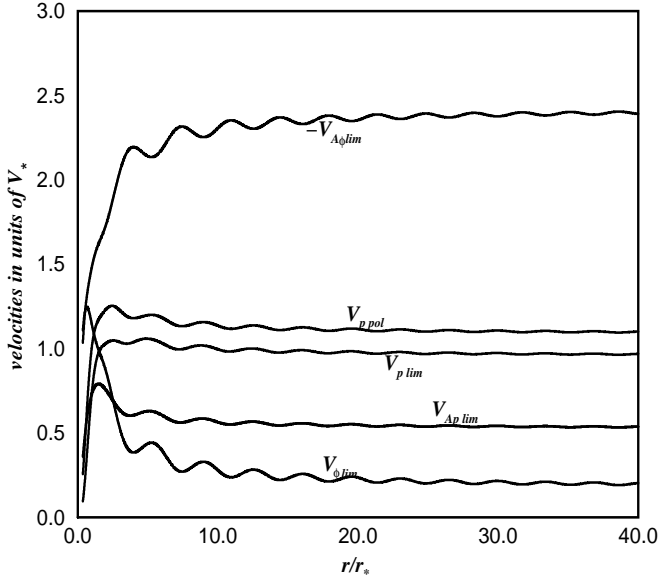,height=9.0truecm,width=10.0truecm,angle=360}}
\caption{Outflow velocities in units of $V_\star$, the radial speed 
at the Alfv\'en point $(\alpha=0,R=1)$, for the parameters given in the 
caption of Fig. (1) of model (2) of Table 1.}
\end{figure}
The oscillatory structure of all flow speeds before the flow reaches full 
cylindrical collimation is also shown in Fig. (3) where 
we have plotted the characteristic velocities in units of the Alfv\'en 
speed at the polar axis and Alfv\'en sphere ($\alpha=0\,,R=1$), $V_\star$. 

The poloidal speed along the polar axis $V_{p, pol}$ increases to 
a uniform superAlfv\'enic value and is higher than the same speed along 
the limiting streamline $V_{p, lim}$ (i.e., the last fieldline rooted 
on the stellar base $r_{star}$ taken to be at $0.315 r_{\star}$). 
Both reach asymptotically uniform values after $V_{p, lim}$ intersects 
the curve of the poloidal Alfv\'en speed $V_{Ap, lim}$ at $R=1$. 
Note that corotation may be seen up to the Alfv\'en distance $R=1$: 
the azimuthal speed $V_{\phi, lim}$ at the 'limiting fieldine' 
increases until the Alfv\'en surface is reached and drops from angular 
momentum conservation as the outflow expands almost conically. Further away 
however, this speed too levels off to a constant value when full collimation 
is achieved, as expected. Finally, the fact that the jet has a large 
component of toroidal field is reflected by the large values of the 
Alfv\'en speed associatd with the toroidal magnetic field, 
$V_{A \phi, lim}$, as compared to the rotational speed 
$V_{\phi, lim}$.     

\section{Systematic construction of classes of radially selfsimilar 
MHD outflows} 

To construct general classes of radially self-similar solutions, we make 
the following two key assumptions: \\
\noindent
(i) the Alfv\'en Mach number $M_{}^{}$ is solely a function of $\theta$,
\begin{equation}\label{assumptions1r}
M = M(\theta)\,,\qquad M(\theta_\star)=1
\,,
\end{equation}
and\\
(ii) that the poloidal velocity and magnetic fields have a dipolar 
angular dependence,
\begin{equation}\label{assumptions2r}
A= \frac{B_o \varpi_o^2}{2}{\cal A}\left(\alpha\right)\,, \qquad
\alpha=\frac{R^2}{G^2\left(\theta\right)}\sin^2 \theta\,,
\quad R={r \over \varpi_o}
\,,
\end{equation}
where $B_o,\varpi_o$ are constants. 
By choosing $G\left(\theta_\star\right)=1$ at the Alfv\'en transition $\theta_\star$, 
$G(\theta)$ evidently measures the cylindrical distance $\varpi$ to the 
polar axis of each fieldline labeled by $\alpha$, normalized to its 
cylindrical distance $\varpi_{\alpha}$ at the Alfv\'en point, 
$G\left(\theta\right)={\varpi}/{\varpi_{\alpha}}$. For a smooth crossing of 
the Alfv\'en cone $\theta=\theta_\star$ [$r=r_{\alpha}\left(\alpha \right)
, \theta = \theta_\star$], 
the free integrals $L$ and $\Omega$ are related by 
\begin{equation}\label{}
 {L\over \Omega } = 
\varpi_{\alpha}^2 (A) = r_{\alpha}^2 \left(\alpha \right) \sin^2 \theta_\star 
=\varpi_o^2 \alpha \,.
\end{equation}
Therefore, the second assumption is equivalent with the statement that  
at the Alfv\'en conical surface, the cylindrical distance $\varpi_a$ of each 
magnetic flux surface $\alpha=const$ is simply proportional to $\sqrt{\alpha}$, 
exactly as in the previous meridionally self-similar case. 
\\
Instead of using the three functions of $\alpha$, (${\cal A}\,, 
\Psi_A$, $\Omega$) we found it more convenient to work with the 
three dimensionless functions of $\alpha$, ($q_1\,,q_2\,,q_3$),  
\begin{equation}\label{q1}
q_1\left( \alpha \right)= \int {{\cal A}^{'2}\over \alpha } d\alpha
\,,
\end{equation}
\begin{equation}\label{q2}
q_2 \left( \alpha \right)=\frac{\varpi_o^2}{B_o^2} \int \Omega ^2 \Psi_A ^2 
d \alpha
\,,
\end{equation}
\begin{equation}\label{q3}
q_3\left(\alpha \right)=\frac{{\cal GM }}{B_o^2 \varpi_o} 
\int \frac{\Psi_A ^2}{\alpha^{{3\over 2}}} d\alpha 
\,. 
\end{equation}
Following the same algorithm as in the previous case, we shall use 
($\alpha\,, \theta$) as the independent 
variables and transform the derivatives with respect to $r$ and 
$\theta$ to derivatives with respect to $\alpha$ and $\theta$ in the 
$\hat r$- and $\hat \theta$-components of the momentum equation. 
Integrating the resulting $\hat r$--component of the momentum equation
we get
\begin{equation}\label{prer}
P=\frac{B_o^2}{8 \pi} \left(h_1 \alpha q_{1}^{'}+h_2 \alpha q_{2}^{'}+h_3 q_2+ 
h_4 q_3 + h_5 q_1+ h_0 \right)
\,,
\end{equation}
or \[P=\frac{B_{o}^2}{8 \pi} \bf{Y} \bf{P}^{\dag} \] with
\begin{equation}\label{prernew}
{\bf{P}}=\left[ \; h_0 \; h_5 \; h_1 \; h_3 \; h_2 \; h_4 \; 0 \; \right]
\,,
\end{equation}
and
\begin{eqnarray}\label{Yrinew}
\begin{array}{l}
{\bf{Y}}=\left[\;Y_1\;Y_2\;Y_3\;Y_4\;Y_5\;Y_6\;Y_7\;\right]=
\\ \\
\left[\;1\;q_1\;\alpha q_1^{'}\;q_{2}\;\alpha q_{2}^{'}\;q_3\;
\alpha q_{3}^{'}\;\right]
\,,
\end{array}
\end{eqnarray}
and after substituting the pressure in the other component 
of the momentum equation we obtain  
\begin{eqnarray}\label{athrr}
\begin{array}{l}
H h_4\alpha q_3^{'}+ h_4^{'}q_3 +h_3 \left(H-2\right)\alpha q_2^{'}+
h_3^{'}q_2+ \\ \\
\frac{\left(h_1 \left[1-M^2\right]^2\right)^{'}\alpha q_{1}^{'}}{
\left(1-M^2\right)} + h_5^{'}q_1 + h_0^{'}=0
\,,
\end{array}
\end{eqnarray}
where a prime in the functions of $q_i(\alpha)$, i=1,2,3 and 
$h_i$ indicates a derivative with respect to their variables 
$\alpha$ and $\ln \sin\theta $, respectively, while 
the functions $h_j\left(\theta\right),j=1,2,3,4,5$ and $H$ 
are given in Appendix B.\\
This expression is again of the form \\
$X_7\left(\theta\right)Y_7\left(\alpha\right)+X_{6}\left(\theta\right)
Y_{6}\left(\alpha\right)+\cdots +X_{1}\left(\theta\right)
Y_1\left(\alpha\right)=0 \,,$ 
\begin{equation}\label{7-1r}
\mbox{or} \,,
{\bf{Y} \bf{X}}^{\dag} = {\bf{0}}
\end{equation}
with {\bf X} the (1 $\times$ 7) matrix
\begin{eqnarray}\label{Xirnew}
\begin{array}{l}
{\bf{X}}=\left[\;X_1\;X_2\;X_3\;X_4\;X_5\;X_6\;X_7\;\right]=
\\ \\
\left[h_0^{'}\;h_5^{'}\;\frac{\left(h_1 \left[1-M^2\right]^2\right)^{'}}
{\left(1-M^2\right)}\;h_3^{'}\;h_3\left(H-2\right)\;h_4^{'}\;H h_4\; \right] 
\,.
\end{array}
\end{eqnarray}

As in the previous case of meridionally selfsimilar solutions, we classify 
the various possibilities by the sets $xxxxxxx$. And, these sets 
always have "00" at the end, their first digit is "1", they have at most 
three "0's", while from the $2^7$ possibilities we end up again with the 
cases 
1011100, 1101100, 1110100, 1111000, 1111100. 
Now the vectors
$q_{1}(\alpha),\,q_{2}(\alpha),\,q_{3}(\alpha)$ belong to a 3D $\alpha$-space 
with basis vectors [$ e_1(\alpha)\,, e_2(\alpha)\,, 
 e_3(\alpha)$]. 
This space contains all vectors $q_i(\alpha)$, $i=1,2,3$ subject to the 
$r$-selfsimilarity constraint manifested by Eq. (\ref{athrr}), i.e.,  
that for a given such set $q_i(\alpha)$, $i=1,2,3$, the vectors 
$1,\alpha q_{i}^{'}(\alpha)$ , $i=1,2,3$ also belong to 
the same space. 
Each of the functions $q_i(\alpha)$, $i=1,2,3$ which satisfy this constraint 
are then a linear combination of the basis vectors $ e_1(\alpha), 
 e_2(\alpha),  e_3(\alpha)$.
In the following, we choose $ e_1=1$, $ e_2= q_1(\alpha )$. 
All such sets of basis vectors give all possible radially 
selfsimilar solutions. Therefore, collecting all possibilities, we 
end up with the 6 classes of solutions shown in Table 3.
\begin{table*}
\centering
\begin{minipage}{140mm}
\caption{Radially Selfsimilar Models}
\begin{tabular}{@{}cllll@{}}
Case& $q_1(\alpha)$ & $q_2(\alpha)$ & $q_3(\alpha)$ & constants \\
\hline
(1) & $\frac{E_1}{F-2} \alpha^{F-2}$ 
& $\frac{D_1}{F-2} \alpha^{F-2}$ 
& $\frac{C_1}{F-2} \alpha^{F-2}$ & $E_1,F-2 \neq 0$ \\
(2) & $E_1 \ln{\alpha}$
& $D_1 \ln{\alpha}$
& $C_1 \ln{\alpha}$ & $E_1 \neq 0$\\
 (3) & $E_1 \alpha^{x_1}+E_2 \alpha^{x_2}$ 
& $D_1 \alpha^{x_1}+D_2 \alpha^{x_2}$
& $C_1 \alpha^{x_1}+C_2 \alpha^{x_2}$ & 
$E_1^2+D_1^2+C_1^2,E_2,x_1,x_2,x_1-x_2 \neq 0 $\\ 
(4) & $E_1 \ln{\alpha} +E_2\alpha^{x}$
& $D_1 \ln{\alpha} +D_2\alpha^{x}$
& $C_1 \ln{\alpha}+C_2\alpha^{x}$ & 
$E_i^2+D_i^2+C_i^2,x \neq 0,i=1,2 $
\\
(5) & $E_1 \left( \ln{\alpha} \right) ^2 +E_2 \ln{\alpha}$
& $D_1 \left( \ln{\alpha} \right) ^2 +D_2 \ln{\alpha}$
& $C_1 \left( \ln{\alpha} \right) ^2 +C_2 \ln{\alpha}$ &
$E_1^2+D_1^2+C_1^2 \neq 0 $ \\
(6) & $E_1 \alpha^{x} \ln{\alpha}+E_2 \alpha^{x}$ 
& $D_1 \alpha^{x} \ln{\alpha}+D_2 \alpha^{x}$ 
& $C_1 \alpha^{x} \ln{\alpha}+C_2 \alpha^{x}$ &
$E_1^2+D_1^2+C_1^2 \neq 0 $ \\
\hline
\end{tabular}
\end{minipage}
\end{table*}

In all of the cases of Table 3, from Eqs. ({\ref{q1}}), ({\ref{q2}}), 
({\ref{q3}}) we find  the form of the functions of $\alpha$,\\
\begin{eqnarray}\label{int}
& & A=\frac{B_o \varpi_o^2}{2} \int_{0}^{\alpha} 
\sqrt{\alpha q_1^{'}} d\alpha\,,\qquad 
\Psi_{A}^2=\frac{B_o^2 \varpi_o }{{\cal GM }} \alpha^{{3 \over 2 }} q_3^{'}\,,
\nonumber \\ \nonumber \\
& & 
\Omega^2=\frac{{\cal GM}}{\varpi_o^3} \frac{q_2^{'}}{q_3^{'}} 
\alpha^{-{3 \over 2}}\,,\qquad 
L^2={\cal GM}\varpi_o\frac{q_2^{'}}{q_3^{'}} \alpha^{{1 \over 2}} 
\,.
\end{eqnarray}
Finally, by substituting $q_1\,,q_2\,,q_3$ in 
Eqs. ({\ref{prer}}) , ({\ref{athrr}}),
we find the ordinary differential equations which the functions 
$G(\theta)\,,M(\theta)\,,h_0(\theta)$ obey.\\
In $\alpha$-space, for each of the cases of Table 3 
there exists a (3 $\times$ 7) matrix ${\bf {K}}$ such that 
\begin{equation}{}
{\bf {Y}}=\left[\;{ e_1}\,\;{ e_2}\,\;{ e_3}\;\right]\;\;{\bf {K}}
\,,
\end{equation}
and from  Eq. ({\ref{7-1r}}) 
\[
\left[\;{ e_1}\,\;{ e_2}\,\;{e_3}\,
\right]\;{\bf{K}}  {\bf{X}}^{\dag}={\bf{0}} \,.\] 
If the basis vectors ${ e_i}$ are linearly independent, then, 
\[
{\bf{K}}  {\bf{X}}^{\dag}={\bf{0}} \,.\]
These three equations are the \underline{ordinary} differential equations for the 
functions of $\theta$ in each model of Table 3, while for the pressure,
\[
P=\frac{B_{o}^2}{8 \pi} \left[\;{e_1}\,\;{e_2}\,\;{e_3}
\,\right] {\bf{K}} {\bf{P}} ^{\dag} 
= \frac{B_{\star}^2}{8 \pi} \left( P_0 + P_1q_1+P_2  e_3 \right)
\,, \] 
where
\[
{\bf K} {\bf P}^{\dag}= [P_0 \; P_1 \; P_2]^{\dag} 
\,.
\]
As with the previous meridionally selfsimilar solutions, the first 
two classes are of particular interest. 
The first corresponds to the following form of the free integrals:
\begin{eqnarray}\label{conto}
A=\frac{B_o \varpi_o^2 \sqrt{E_1}}{F}\alpha ^{{F \over 2}}\,,\;
\Psi_A^2=\frac{C_1B_o^2 \varpi_o}{{\cal GM}} \alpha^{(F-3/2)}\,,
\nonumber \\
\Omega^2=\frac{D_1{\cal GM}}{\varpi_o^3 C_1}\alpha^{-{3\over 2}}
\,.
\end{eqnarray}

This is a degenerate case, i.e., 
$ e_3 = 0$ and we have only two conditions between 
the functions of $\theta$. It follows that we are free to impose a third 
relation between the unknown functions $[G(\theta)\,, M(\theta)\,, 
h_0(\theta)]$. One possibility is that such a third imposed relation 
is of the polytropic type, $P \propto \rho^{\gamma}$ (in this case $h_0=0$). 
In such a polytropic case which has been analysed in detail in Contopoulos 
\& Lovelace (1995), the magnetic flux is of the form $A = f_f(\theta) R^F$ with 
$f_f(\theta) \propto [\sin \theta /G(\theta )]^{F}$ (for notation see also 
Tsinganos et al 1996). The magnetic field at the equatorial plane 
$\theta=90^o$ is $B\propto R^{F-2}$, the density $\rho \propto 
R^{2F-3}$, while the sound, Alfv\'en and rotational speeds scale as their 
Keplerian counterparts, i.e., as $R^{-1/2}$. Note that if 
$[{D_1 G\left({\pi}/{2} \right) }/{C_1}] \left[
{(G^2-M^2)}/{G \left(1-M^2\right)}\right]_{\theta= \frac{\pi}{2}} ^2=1$,  
the rotational velocity at the equatorial plane is exactly Keplerian.
The classical and simplest subcase analysed in BP82 corresponds to 
the subclass with $F=3/4$, wherein $B\propto R^{-5/4}$.   
The two relations among the functions of $\theta$ are the two 
resulting first order differential equations for the Alfv\'en number 
$M(\theta)$ and dimensionless radius $G(\theta )$ [$m(\chi)=M^2(\theta)$ and 
$\xi (\chi )={G(\theta)}/{G(\frac{\pi}{2})}$ in the notation of BP82].\\
The second case is also degenerate since 
$ e_3 = 0$ with again only two conditions between 
the functions of $\theta$. As before, we are free to impose a third 
relation between the unknown functions $[G(\theta)\,, M(\theta)\,, 
h_0(\theta)]$, for example, a polytropic relationship. 
Then one can prove that this case is a subcase of the first one 
(if it is polytropic), for $F=2$.
All other cases shown in Table 3 are nondegenerate.

The third class, is characterized {\it first} by a set of parameters 
describing the particular model and the dependance of the free integrals 
on the magnetic flux function $A\left(\alpha \right)$, ($x_1\,, x_2\,, 
E_1\,, E_2\,, C_1\,, C_2\,, D_1\,, D_2)$. 
{\it Second} by the Alfv\'en angle $\theta_\star$.
And {\it third}, by the set of the critical point  
parameters $p_\star = \left( {dM^2}/{d\theta}\right)_\star$ and $\varphi_{\star}$ which 
denote the slope of the Alfv\'en number and the expansion angle, respectively, 
at the Alfv\'en angle $\theta_\star$, together with the pressure component 
$P_{1\star}$ through $h_{5\star}$. This triplet of 'dynamical' parameters fixes the physical 
solution and they are related through the Alfv\'en regularity condition which 
is now obtained from Eq. (\ref{h5new}) of Appendix B at the Alfv\'en angle 
$\theta_\star$ where $M=G=1$ and $h_5=h_{5\star}$, i.e., 
\,.
\begin{equation}\label{r-Alfven}
h_{5\star}=-{\sin^2\theta_{\star} \tan(\theta_\star + \varphi_\star )} p_\star
\,.
\end{equation}
As with the previous case of meridional selfsimilarity, this condition relates the 
slope of the square of the Alfv\'en number $p_\star =
\left( {dM^2}/{d\theta}\right)_\star$ and the expansion angle $\varphi_\star$ 
with the pressure component 
$P_{1\star}$ 
through $h_{5\star}$. 
Finally, the requirement that the solution crosses the two 
slow and fast X-type critical points (modified by the radial self-similarity 
assumption, Tsinganos et al 1996) determines all these three 'dynamical' parameters
$[\varphi_\star\,, p_\star\,, P_{1\star} ]$.
 
It is interesting to note that contrary to classes (1)-(2) in Table 3, 
this model (3) may be characterized by a scale, for example the radial 
distance on the plane of the disk where the magnitudes of the poloidal 
speed and magnetic field or the toroidal speed and magnetic field become zero. 
Hence, it occured to us that this is an interesting generalisation of the 
BP82 model and therefore worthy of further investigation. 

\begin{figure}
\centerline{\psfig{file=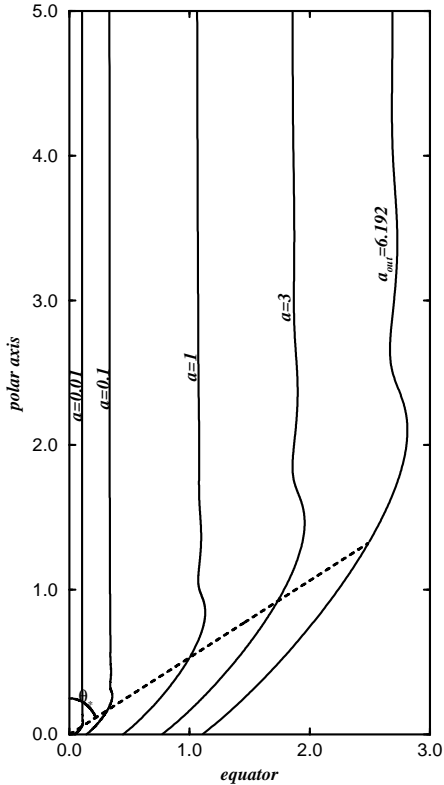,height=13.0truecm,angle=360}}
\caption{Field and streamlines for the cylindrical $r$-self similar model 
of case (3) from Table 3 and the following set of parameters: 
$x_1=-0.6$, $x_2=-0.5$, 
$E_1=-0.03$, $E_2=0.03$, $C_1=-1.5$, $C_2=-0.6$, $D_1=-25$, 
$D_2=-10$, $\theta_\star=62^o$, $\varphi_\star=55^o$, 
$p_\star=-3$. At the disk level, $V_{\phi}\propto R^{-1/2}$ 
while on the poloidal field/streamline $\alpha_{out}=6.191736$, $B_{p}= V_{p}=0$.}
\end{figure}

\begin{figure}
\centerline{\psfig{file=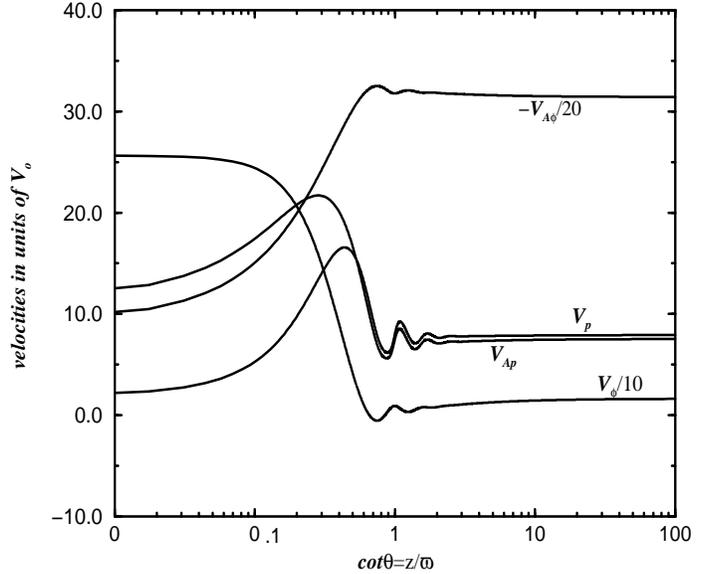,height=9.0truecm,width=10.0truecm,angle=360}}
\caption{The characteristic velocities of model (3) of Table 3 with 
cylindrical asymptotics are plotted in units of the z-component of the 
flow speed at the point $(\alpha=1,\theta=\pi/2)\,,V_o$ and the same parameters 
as in Fig. (4).}
\end{figure}

Figs. (4-5) are a typical illustration of model (3) for describing 
collimated jet-type outflows \underbar{with} an oscillatory behaviour. In 
Fig. (4) the poloidal field and streamlines reach a cylindrical shape after 
undergoing oscillations in their radius. As we move downstream, the amplitude 
of these oscillations decays while their wavelength increases. In fact, the exact 
behaviour of the oscillations is analytically described in Vlahakis \& 
Tsinganos (1997) where it is shown that they can be regarded as perturbations on 
an asymptotically cylindrical shape which can be expressed in terms of the Legendre 
functions $P_{\nu}^{\mu}(\cos\theta)$ and $Q_{\nu}^{\mu}(\cos\theta)$. 
According to this analysis, when 
$\mu^2 <0$, the asymptotically cylindrical shape is finally obtained through 
those oscillations. Then the perturbation (for $\theta \rightarrow 0$)
is proportional to $\theta ^{\pm \mu-\nu}$, or 
since $\mu^2 <0$, proportional to $\left(\frac{\varpi}{z}\right)^{-\nu}
\cos \left( \mid {\mu} \mid \ln \frac{\varpi}{z} + D_o \right)$.
In the example shown in Fig. (4-5) the amplitude of the 
oscillations is rather weak. Note however, that cases also exist with an 
extremely strong oscillation amplitude and such examples will be analysed in 
another connection. 
On the other hand, when $\mu^2 \geq 0$ the asymptotically cylindrical shape is 
reached without such oscillations. Exactly this last possibility is shown  
in the following case of Figs. (6-7). 

\begin{figure}
\centerline{\psfig{file=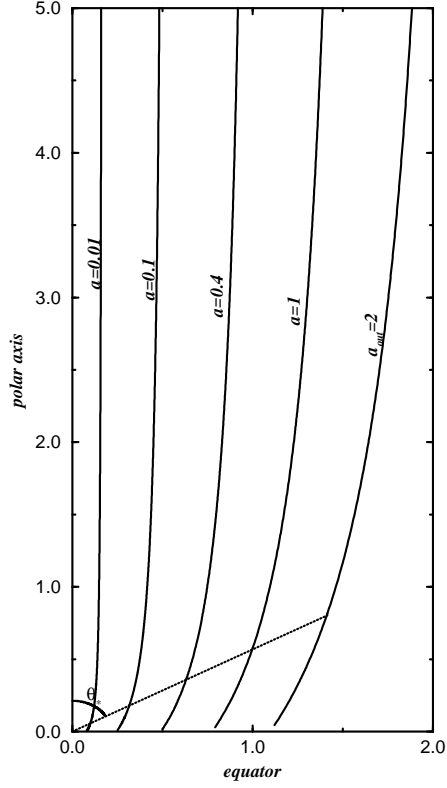,height=13.0truecm,angle=360}}
\caption{Field and streamlines for the cylindrical $r$-self similar model 
of case (3) from Table 3 and the following set of parameters: 
$x_1=-0.9$, $x_2=-0.6$, 
$E_1=-2.1421466$, $E_2=2.60994552$, $C_1=-3.2132198$, $C_2=D_2=0$, 
$D_1=-160.66099$, $\theta_\star=60^o$, $\varphi_\star=74.704656^o$, 
$p_\star=-1.1$. At the disk level, $V_{\phi}\propto R^{-1/2}$ 
while on the poloidal field/streamline $\alpha_{out}=2$, $B_{p}= V_{p}=0$.}
\end{figure}

\begin{figure}
\centerline{\psfig{file=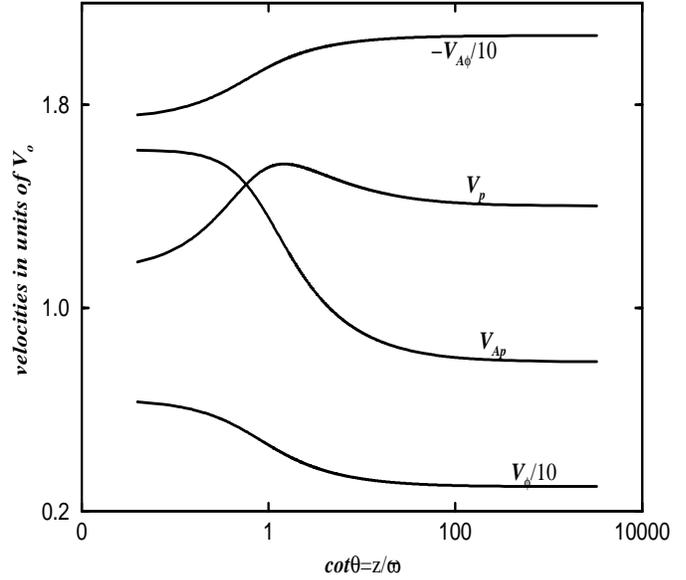,height=9.0truecm,width=10.0truecm,angle=360}}
\caption{The characteristic velocities of model (3) of Table 3 with 
cylindrical asymptotics are plotted in units of the z-component of the 
flow speed at the point $(\alpha=1,\theta=\pi/2)\,,V_o$ and the same parameters 
as in Fig. (6).}
\end{figure}

To further illustrate the various possibilities for the asymptotic behaviour 
of outflows starting from a Keplerian disk, we examine briefly the group 
of three models in 
Figs. (6-7), (8-9) and (10-11) where depending on the values of the model constants, 
we get one with cylindrical, parabolical, or conical terminal geometry: 

(1) In Figs. (6-7) a {\it cylindrically} collimated outflow 
(when $\theta \rightarrow 0 \,, (M^2\,, G^2) \rightarrow \; constants$)
is obtained for 
a set of the model parameters: ($x_i$, $E_i$, $C_i$, $D_i$), 
i=1,2. The Alfv\'en conical 
surface is taken at $\theta_\star=60^o$ where the slope of the square of the 
Alfv\'en number is fixed as $p_\star=-1.1$ while the expansion angle   
$\varphi_\star\approx 75^o$ (the angle of the poloidal streamline with the 
cylindrical radius). The characteristic scale of the model is taken 
to indicate approximately the  
radius of the jet, or more precisely, the distance along the disk where 
for $\alpha_{out}=2$ we have $B_{p}=V_{p}=0$.
In Fig. (7) the velocities on the reference line $\alpha=1$ 
are plotted in units of $V_o$, the z-component 
of the flow speed at the point $(\alpha=1,\theta=\pi/2)$. 

\begin{figure}
\centerline{\psfig{file=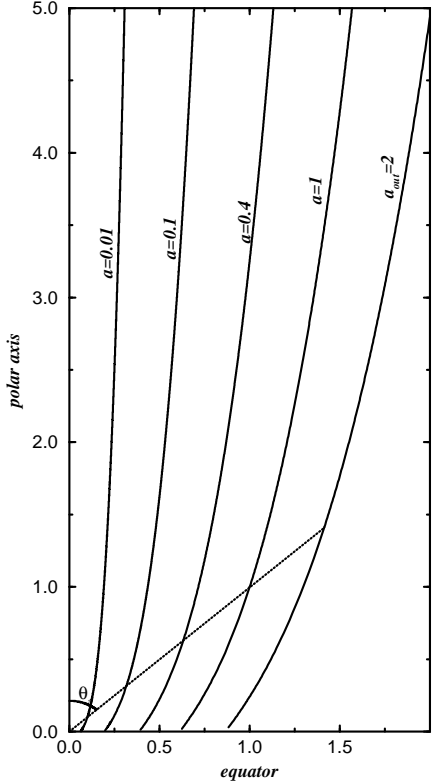,height=13.0truecm,angle=360}}
\caption{Poloidal field and streamlines for the parabolic $r$-self similar model 
of case (3), Table 3 and the following set of parameters: 
$x_1=-0.9$, $x_2=-0.6$, $E_1=-0.8252542$, $E_2=1.00547$, 
$C_1=-1.23788$, $C_2=D_2=0$, $D_1=-12.378813$, $\theta_\star=45^o$, 
$\varphi_\star=75.465545^o$, $p_\star=-1.7$.
In this case $V_{\phi}\propto R^{-1/2}$ on the equatorial plane while 
on the streamline $\alpha_{out}=2$, $B_{p}=V_{p}=0$.}
\end{figure}

\begin{figure}
\centerline{\psfig{file=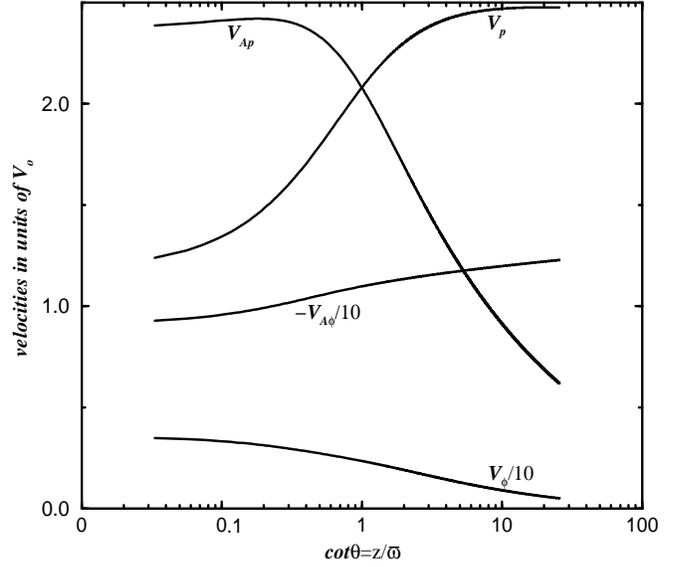,height=9.0truecm,width=10.0truecm,angle=360}}
\caption{The characteristic velocities of model (3) of Table 3 with 
paraboloidal asymptotics are plotted in units of the z-component of the flow 
speed at the point $(\alpha=1,\theta=\pi/2)\,,V_o$ and the same parameters as in 
Fig. (8).}
\end{figure}

(2) In Figs. (8- 9) an $r$-self similar model belonging to case (3) in Table 3 
with {\it parabolic} asymptotical geometry 
(when $\theta \rightarrow 0 \,, (M^2\,, G^2) \rightarrow \infty$)
is examined for another set of 
parameters ($x_i$, $E_i$, $C_i$, $D_i$), i=1,2. The Alfv\'en conical 
surface is taken now at $\theta_\star=45^o$ where the slope of the square of the 
Alfv\'en number is chosen as $p_\star=-1.7$ and the expansion angle 
$\varphi_{\star} \approx 75^o$. 

\begin{figure}
\centerline{\psfig{file=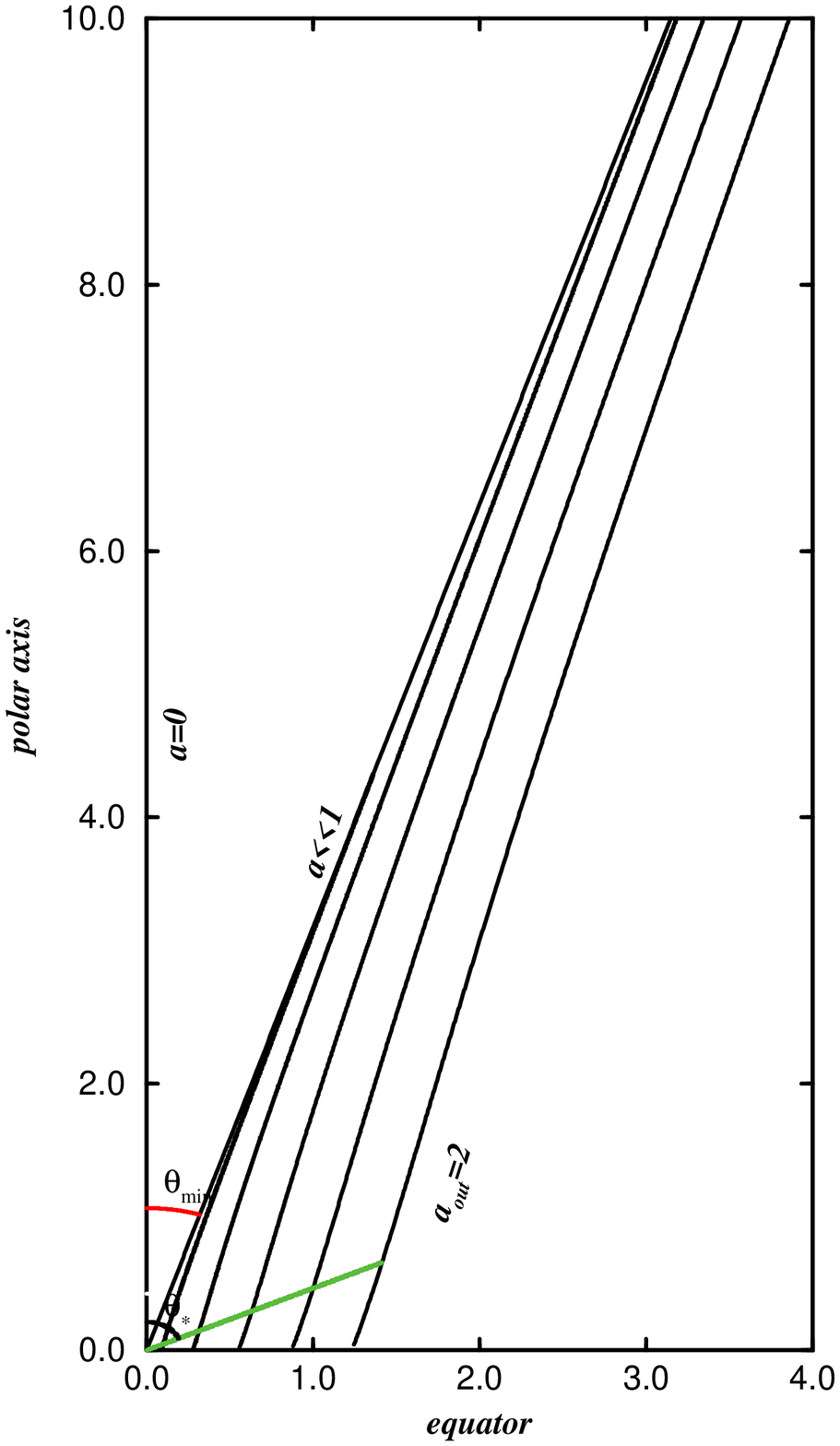,height=13.0truecm,angle=360}}
\caption{Field and streamlines for the conical $r$-self similar model 
of case (3), Table 3 and the following set of parameters: 
$x_1=-0.1,x_2=0.01$ and $E_1=-78.601635$, $E_2=-728.31337$,
$C_1=-4.3231$, $C_2=D_2=0$, $D_1=-43.231$, $\theta_\star=65^o$, 
$\varphi_\star=75.784234^o$, $p_\star=-0.5$.
In this case $V_{\phi}\propto R^{-1/2}$ on the equatorial plane while 
on the poloidal field/streamline $\alpha_{out}=2$, $B_{p}= V_{p}=0$.
For large distances from the disk all lines with $\alpha>0$ go asymptotically to 
the line $\theta=\theta_{min}$.}
\end{figure}

\begin{figure}
\centerline{\psfig{file=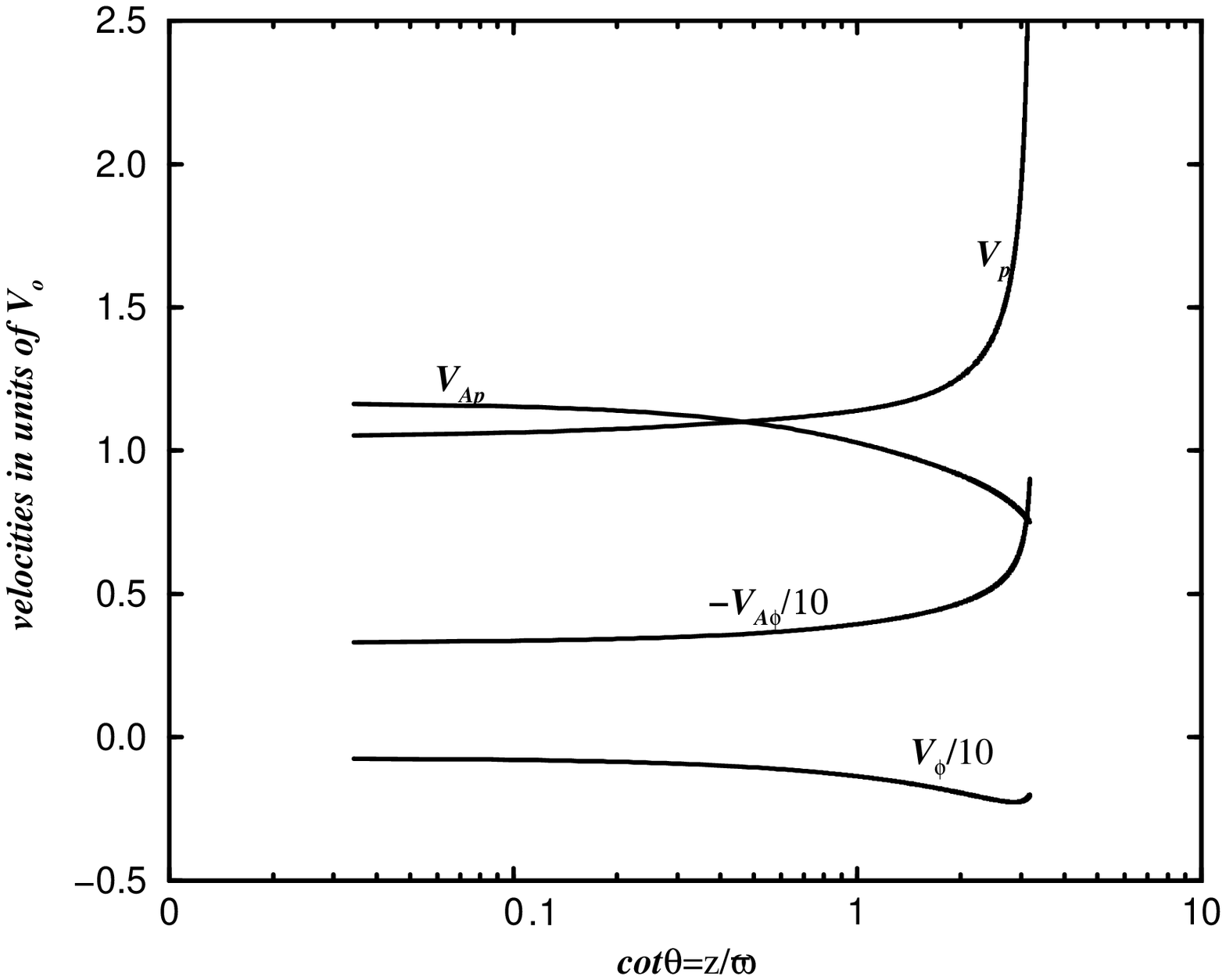,height=9.0truecm,width=10.0truecm,angle=360}}
\caption{The characteristic velocities of model (3) of Table 3 with 
conical asymptotical geometry are plotted in units of the z-component of the flow 
speed at the point $(\alpha=1,\theta=\pi/2)\,,V_o$ and the same parameters as in 
Fig. (10).}
\end{figure}

(3) Finally, in Figs. (10-11) the $r$-self similar model of case (3) in Table 3 
gives a {\it conical} asymptotical geometry for a third set of the parameters 
($x_i$, $E_i$, $C_i$, $D_i$), i=1,2 and $\theta_\star=65^o$, 
$\varphi_\star =75^o$, $p_\star=-0.5$.
Note that now the solution exists only for $\theta > \theta_{min}$ where 
$\theta_{min}\approx 17.5^o$. 
When this value of $\theta$ is approached, 
$(M^2\,, G^2) \rightarrow \infty$.

In all these four possibilities and along 
a given field/streamline, the outflow starts from the equator where 
$V_{\phi}\propto R^{-1/2}$ with a low subAlfv\'enic poloidal speed. 
This poloidal speed $V_{p}$ crosses the Alfv\'en conical surface at 
$\theta_{\star}$ in all cases. In the cylindrical case of Fig. (7),  
$V_{p}$ increases rapidly to a uniform value when collimation is achieved 
. The azimuthal speed $V_{\phi}$ on the other hand, 
drops with height in all cases, as rotational energy is transformed to poloidal 
kinetic energy. Finally, the azimuthal Alfv\'en speed is the strongest 
in the cylindrical case where the toroidal magnetic field is responsible 
for the ensuing final collimation.  

\section{Summary}

In this paper we have examined a systematic way for constructing exact 
MHD solutions for plasma flows. The {\it first} assumption was to 
consider the ideal plasmas MHD equations for time-independent conditions, 
Eq. (\ref{momentum}-\ref{fluxes+F}), {\it without} imposing the extra 
constraint of the frequently used polytropic assumption.  
{\it Second}, we confined our attention to axisymmetric situations in 
which case the poloidal magnetic and velocity fields can be expressed in 
terms of the magnetic flux function $A$ while several integrals exist, 
Eq. (\ref{Bfield}-\ref{Vfield}). In that case, besides $A$, a second  
natural variable is the Alfv\'en Mach number $M$, Eq. (\ref{Alfven}).
We denoted by $G$ the cylindrical distance $\varpi$ of a poloidal streamline 
from the system's symmetry axis, in units of the cylindrical distance of 
the Alfv\'en surface from the same axis, $\varpi_{\alpha}$.
{\it Third}, we further confined our attention to transAlfvenic outflows 
in which case the regularization of the azimuthal components in 
Eq. (\ref{Bfield}-\ref{Vfield}) requires that the ratio of the two integrals 
of the total specific angular momentum in the flow $L(A)$ and corotation 
frequency $\Omega(A)$ is some function $\alpha (A)$ [as in 
Eq. (\ref{regularity_phi})]. 
By introducing some reference scale $\varpi_o$ 
this function $\alpha$ is dimensionless, [as in Eq. (\ref{assumptions2}) where 
$\varpi_o\equiv r_\star$]. 
Apparently ($M, \alpha$) is a rather convenient set of dimensionless variables 
for describing all physical quantities in the poloidal plane. For any set of 
orthogonal curvilinear coordinates suitable for describing axisymmetric problems, 
we may then convert their poloidal coordinates to ($M, \alpha$). Examples are, 
spherical coordinates [$r(M, \alpha), \theta (M, \alpha ), \phi$], 
cylindrical coordinates [$z(M, \alpha), \varpi (M, \alpha ), \phi$], 
toroidal coordinates [$u(M, \alpha), v(M, \alpha ), \phi$], 
oblate/prolate spheroidal coordinates [$\xi(M, \alpha), \eta (M, \alpha ), \phi$], 
paraboloidal coordinates, etc.     
Then, the distance from the symmetry axis of the outflow is $G(M, \alpha)$. 
In the present first study we made the simplifying {\it fourth} assumption 
that $G$ is independent of $\alpha$, $G=G(M)$ only. Finally, to re-establish 
the connection with the geometry of the problem and the particular set of the 
coordinates used, we made our {\it fifth} and final 
assumption that $M=M(\chi)$ 
(and $G=G(\chi)$), where $\chi = r$, or, $\chi = \theta$. This leads then to the 
two broad classes of meridionally and radially self-similar outflows. 
Needless to say that additional symmetries may in principle be considered, 
something which may be taken up in another connection (equilibria in tokamak 
geometries, etc).   

After these five assumptions are well posed and with the help of a simple 
theorem, it is possible to (i) unify all existing exact solutions for 
astrophysical outflows (Tables 1,2 and 3) and (ii), to qualitatively 
sketch a few of them. With this method, the system of the coupled 
MHD equations reduces to a set of five ordinary differential equations 
for the dimensionless jet radius ($G$), the flow's expansion factor or angle 
($F$, or $\varphi$), the Alfv\'en Mach number ($M$) and the two pressure 
components ($P_1$ and $P_0$). The requirement that the solutions pass through 
the Alfv\'en critical point gives a condition relating the values of the 
expansion function or angle, Alfv\'en number slope and pressure 
component at this critical point. 
The Alfve\'n regularity conditions, Eqs. (\ref{r-Alfven}),
(\ref{t-alfven}) is similar to that 
discussed in Heyvarts \& Norman (1989) and ST94.

As a byproduct of this construction, two representative models for radially and 
meridionally self-similar outflows, BP82 and ST94, respectively, have 
been generalized. In the former case of BP82, it is well known that the 
cold plasma solution is terminated at a finite height above the disk while 
the general case (3) in Table 3 extends all the way to infinity. Also, it is 
shown that the expressions of the MHD integrals which correspond to the ST94 
model are only a special case of case (2) in Table 1.  

Having in mind the ubiquitously observed collimated 
outflows from astrophysical objects, we paid more attention to 
the selfconsistently derived asymptotical shape of the streamlines. Of the 
various such asymptotic geometries derived, a prominent member seem to 
be the cylindrically collimated jet-type solutions, in accordance also 
with the conclusions of observations (Livio 1997), general theoretical 
arguments (Heyvaerts \& Norman 1989) and recent numerical simulations 
(Goodson et al 1997). Another feature that appeared in the solutions is 
that cylindrical collimation may or may not be achieved with oscillations in 
the width of the jet (Vlahakis \& Tsinganos 1997). Although in the 
examples analyzed here the amplitude of the oscillations is rather weak 
and the flow collimates rather smoothly, preliminary results show that 
cases also exist where it can become rather large and the final radius of 
the jet can be much smaller than the initial large cylindrical radius and 
corresponding opening angle. Finally, we should note that the pressure 
$P$ denotes the total pressure (including gas pressure, Alfv\'en waves 
pressure, radiative forces, etc). For example, the same formalism 
may be used also in radiation driven winds. 

\section*{Acknowledgments}

This research has been supported in part by the grant 107526 of the General 
Secretariat of Research and Technology of Greece. We thank J. Contopoulos, 
C. Sauty and E. Trussoni for helpful discussions.

\appendix
\section[]{functions of $R$} 

\begin{equation}\label{F}
F=2-R\frac{G^{2'}}{G^2}
\equiv  \frac{\partial \ln \alpha (R, \theta) }{\partial \ln R}
\,.
\end{equation}
\begin{equation}\label{f1}
f_1=-\frac{1}{G^4}
\,,
\end{equation}
\begin{equation}\label{f2}
f_2=-\frac{F^2-4}{4G^2 R^2}
\,,
\end{equation}
\begin{equation}\label{f3}
f_3=-\frac{1}{G^2}\left(\frac{1-G^2}{1-M^2}\right)^2
\,,
\end{equation}
\begin{equation}\label{f4}
f_4=\frac{F }{2RG^2}M^{2'}-\frac{1-M^2}{2RG^2}F^{'}-\frac{\left(1-M^2\right)F\left(F-2\right)}{4R^2 G^2}
\,,
\end{equation}
so,
\begin{equation}\label{F'}
F'=\frac{F}{1-M^2}M^{2'}-\frac{F\left(F-2\right)}{2R}-\frac{2RG^2}{1-M^2}f_4
\,,
\end{equation}
\begin{equation}\label{f5}
f_5=\frac{G^4-M^2}{G^2M^2\left(1-M^2\right)}
\,,
\end{equation}
\begin{equation}\label{f6}
f_6=-\frac{2}{G^4}M^{2'}+\frac{2\left(1-M^2\right)\left(F-2\right)}{RG^4}
\,,
\end{equation}
\begin{equation}\label{f7}
f_7=\frac{2}{R^2 G^2} M^{2'}-\frac{\left(1-M^2\right)\left(F-2\right)\left(F+4\right)}{2R^3G^2}-\frac{F}{R}f_4
\,,
\end{equation}
\begin{equation}\label{f8}
f_8=-\frac{F-2}{R} f_5
\,,
\end{equation}
\begin{equation}\label{f9}
f_9=-\frac{\nu^2}{R^2 M^2}
\,.
\end{equation}

\section[]{functions of $\theta$}
\begin{equation}\label{H}
H=2-\frac{G^{2'}}{G^2}\equiv  
\frac{\partial \ln \alpha (R, \theta)}{\partial \ln \sin \theta}
\,,
\end{equation}
\begin{equation}\label{Hfi}
H=-2\frac{\sin\theta \sin \left(\varphi+\theta\right)}{\cos\theta \cos\left(\varphi+\theta\right)}=
2-2\frac{\cos \varphi}{\cos\theta \cos\left(\varphi+\theta\right)}
\,,
\end{equation}
where the expansion angle $\varphi$ is the angle between the line and 
the equatorial plane, which is a function of $\theta$.
\begin{equation}\label{parG}
\frac{dG^2}{d\theta}=\frac{2 G^2 \cos \varphi}{\sin\theta \cos\left(\varphi+\theta\right)}
\,,
\end{equation}
\begin{equation}\label{h1}
h_1=-{\left(\sin^2 \theta +\cos ^2 \theta {H^2 \over 4} \right) \over G^4}=
-\frac{\sin^2 \theta}{G^4 \cos^2 \left(\varphi+\theta\right)}
\,,
\end{equation}
\begin{equation}\label{h2}
h_2=-\frac{1}{G^2}\left(\frac{1-G^2}{1-M^2}\right)^2
\,,
\end{equation}
\begin{equation}\label{h3}
h_3=\frac{G^4-M^2}{M^2 G^2 \left(1-M^2\right)}
\,,
\end{equation}
\begin{equation}\label{h4}
h_4=-\frac{\sin \theta}{GM^2}
\,,
\end{equation}
\begin{eqnarray}\label{h5}
h_5=-\frac{\cos ^2 \theta}{2 G^4} \left\{
\left[H\left(1-M^2\right)\right]^{'}+\right.
\nonumber \\ \left.
\left(H-2\right) 
\left(1-M^2\right)\left(H-\tan^2 \theta\right)\right\}
\,,
\end{eqnarray}
or,
\begin{eqnarray}\label{h5new}
h_5=\frac{1-M^2}{G^4}\frac{\sin^2 \theta}{\cos^2 \left(\varphi+\theta\right)}
\frac{d\varphi}{d\theta}-\frac{\sin^2 \theta \sin\left(\varphi+\theta\right)}
{G^4 \cos\left(\varphi+\theta\right) }\frac{dM^2}{d\theta}-
\nonumber \\
\frac{1-M^2}{G^4}\frac{\sin\theta \cos \varphi \sin\left(\varphi+\theta\right)}
{\cos^2 \left(\varphi+\theta\right)}
\,,
\end{eqnarray}
so,
\begin{eqnarray}\label{dfi}
\frac{d\varphi}{d\theta}=\frac{\sin\left(\varphi+\theta\right) 
\cos \left(\varphi+\theta\right)}{1-M^2}
\frac{dM^2}{d\theta}+\frac{\sin\left(\varphi+\theta\right) \cos \varphi}{\sin\theta}+
\nonumber \\
\frac{\cos^2 \left(\varphi+\theta\right)}{\sin^2\theta}\frac{G^4}{1-M^2}h_5
\,.
\end{eqnarray}
 
\bsp

\label{lastpage}


\begin{thebibliography}{}
 
   \bibitem{} Bardeen, J.M., Berger, B.K., 1978, ApJ, 221, 105

   \bibitem{} Biretta, T., 1996, in {\it Solar and Astrophysical MHD Flows}, 
      K. Tsinganos (ed.), Kluwer Academic Publishers, 357

   \bibitem{} Blandford, R.D., Payne, D.G., 1982, MNRAS, 199, 883 (BP82)

   \bibitem{} Blandford, R.D., Rees, M.J., 1974, MNRAS, 169, 395

   \bibitem{} Burderi, L., King, A.R., 1995, MNRAS, 276, 1141 

   \bibitem{} Burrows, C.J., et al, 1995, ApJ, 452, 680 

   \bibitem{} Cao, X., 1997, MNRAS, 291, 145 

   \bibitem{} Contopoulos, J., Lovelace, R.V.E., 1994, ApJ., 429, 139 

   \bibitem{} Crane, P., Vernet, J., 1997, ApJ, 486, L91

   \bibitem{} Feldman, W.C., Phillips, J.L., Barraclough, B.L., Hammond, 
      C.M. 1996, in {\it Solar and Astrophysical MHD Flows}, 
      K. Tsinganos (ed.), Kluwer Academic Publishers, 265

   \bibitem{} Ferrari, A., Massaglia, S., Bodo, G., Rossi, P., 
      1996, in {\it Solar and Astrophysical MHD Flows}, 
      K. Tsinganos (ed.), Kluwer Academic Publishers, 607

   \bibitem{} Ferreira, J., 1997, A\&A, 319, 340

   \bibitem{} Ferreira, J., \& Pelletier, G., 1995, A\&A, 295, 807

   \bibitem{} Goodson, A.P., Winglee, R.M., Bohm, K.H., 1997, ApJ, 489, 199

   \bibitem{} Heyvaerts, J., Norman, C.A., 1989, ApJ, 347, 1055

   \bibitem{} Kafatos, M., 1996, in {\it Solar and Astrophysical MHD Flows}, 
      K. Tsinganos (ed.), Kluwer Academic Publishers, 585

   \bibitem{} Konigl, A., 1989, ApJ, 342, 208

   \bibitem{} Li, Z.-Y, 1995, ApJ, 444, 848

    \bibitem{} Lima, J., Tsinganos, K., Priest, E., 1996, Astrophys. Letts. and  
        Comms., 34, 281

   \bibitem{} Livio, M., 1997, in {\it Accretion Phenomena and Related Outflows}, 
       D.T. Wickramasinghe, L. Ferrario, \& G.V. Bicknel (eds.), ASP: San Francisco, 
       845.

   \bibitem{} Mirabel, I.F., Rodriguez, L.F., 1996, in {\it Solar and 
      Astrophysical MHD Flows}, K. Tsinganos (ed.), Kluwer Academic Publishers, 683

    \bibitem{} Parker, E.N. 1958, ApJ, 128, 664

   \bibitem{} Pelletier, G., Pudritz, R.E., 1992,  ApJ, 394, 117

   \bibitem{} Ray, T.P., 1996, in {\it Solar and Astrophysical MHD Flows}, 
      K. Tsinganos (ed.), Kluwer Academic Publishers, 539

   \bibitem{} Sauty, C., Tsinganos, K., 1994, A\&A, 287, 893 (ST94)

   \bibitem{} Trussoni, Tsinganos, K., E.,  Sauty, C., 1997, A\&A, 325, 1099

   \bibitem{} Tsinganos, K.C., 1982, ApJ, 252, 775

   \bibitem{} Tsinganos, K., Trussoni, E., 1991, A\&A, 249, 156

   \bibitem{} Tsinganos, K., Sauty, C., Surlantzis, G., Trussoni, E., 
       Contopoulos, J., 1996, MNRAS, 283, 811

   \bibitem{} Vlahakis, N., Tsinganos, K., 1997, MNRAS, 292, 591

   \bibitem{} Weber, E.J., Davis, L.J. 1967, ApJ, 148, 217

\end{thebibliography}
\end{document}